\documentclass[]{aa}
\usepackage{epsfig}
\usepackage{natbib}
\bibpunct{(}{)}{;}{a}{}{,}

\newcommand{\FeoH}{\ensuremath{\left[\mathrm{Fe}/\mathrm{H}\right]}}
\newcommand{\MoH}{\ensuremath{\left[\mathrm{M}/\mathrm{H}\right]}}
\newcommand{\aoFe}{\ensuremath{\left[\mathrm{\alpha}/\mathrm{Fe}\right]}}

\begin{document}

\title{Broad-band photometric colors and effective temperature calibrations for late-type giants. II. Z$<$0.02}

\author { A. Ku\v{c}inskas \inst{1,2,3}, P.H. Hauschildt \inst{4}, I. Brott \inst{4,5},
        V. Vansevi\v{c}ius \inst{6} \\L. Lindegren \inst{1}, T. Tanab\'{e} \inst{7}, F. Allard \inst{8}}
\institute{
           Lund Observatory, Lund University, Box 43, SE-221 00, Lund, Sweden
    \and   National Astronomical Observatory of Japan, Mitaka, Tokyo, 181-8588, Japan
    \and   Institute of Theoretical Physics and Astronomy, Go\v {s}tauto 12, Vilnius 01108, Lithuania
    \and   Hamburger Sternwarte, Gojenbergsweg 112, 21029 Hamburg, Germany
    \and   INTEGRAL Science Data Centre, Chemin d'Ecogia 16, 1290 Versoix, Switzerland
    \and   Institute of Physics, Savanoriu 231, Vilnius 02300, Lithuania
    \and   Institute of Astronomy, The University of Tokyo, Mitaka, Tokyo, 181-0015, Japan
    \and   Centre de Recherche Astronomique de Lyon, \'{E}cole Normale Sup\'erieure, Lyon, Cedex 07, 69364 France
           }

\offprints{A. Ku\v{c}inskas, \email{ak@itpa.lt}}

\date{Received 28 October 2005 / Accepted 7 March 2006}

\titlerunning{Broad-band colors and effective temperature calibrations for late-type giants. II.}

\authorrunning{Ku\v{c}inskas et al.}

\abstract{We investigate the effects of metallicity on the
broad-band photometric colors of late-type giants, and make a
comparison of synthetic colors with observed photometric
properties of late-type giants over a wide range of effective
temperatures ($T_{\rm eff}=3500-4800$\,K) and gravities ($\log
g=0.0-2.5$), at $\MoH=-1.0$ and $-2.0$. The influence of
metallicity on the synthetic photometric colors is small at
effective temperatures above $\sim3800$\,K, but the effects grow
larger at lower $T_{\rm eff}$, due to the changing efficiency of
molecule formation which reduces molecular opacities at lower
$\MoH$. To make a detailed comparison of the synthetic and
observed photometric colors of late type giants in the $T_{\rm
eff}$--color and color--color planes (which is done at two
metallicities, $\MoH=-1.0$ and $-2.0$), we derive a set of new
$T_{\rm eff}$--$\log g$--color relations based on synthetic
photometric colors, at $\MoH=-0.5$, $-1.0$, $-1.5$, and $-2.0$.
These relations are based on the $T_{\rm eff}$--$\log g$ scales
that we derive employing literature data for 152 late-type giants
in 10 Galactic globular clusters (with metallicities of the
individual stars between $\MoH=-0.7$ and $-2.5$), and synthetic
colors produced with the {\tt PHOENIX}, {\tt MARCS} and {\tt
ATLAS} stellar atmosphere codes. Combined with the $T_{\rm
eff}$--$\log g$--color relations at $\MoH=0.0$ \citep{K05}, the
set of new relations covers metallicities $\MoH=0.0\dots-2.0$
($\Delta\MoH=0.5$), effective temperatures $T_{\rm
eff}=3500\dots4800$\,K ($\Delta T_{\rm eff}=100$\,K), and
gravities $\log g=-0.5\dots3.0$. The new $T_{\rm eff}$--$\log
g$--color relations are in good agreement with published $T_{\rm
eff}$--color relations based on observed properties of late-type
giants, both at $\MoH=-1.0$ and $-2.0$. The differences in all
$T_{\rm eff}$--color planes are typically well within
$\sim100$\,K. We find, however, that effective temperatures
predicted by the scales based on synthetic colors tend to be
slightly higher than those resulting from the $T_{\rm eff}$--color
relations based on observations, with the offsets up to $\sim
100$\,K. This is clearly seen both at $\MoH=-1.0$ and $-2.0$,
especially in the $T_{\rm eff}$--$(B-V)$ and $T_{\rm
eff}$--$(V-K)$ planes. The consistency between $T_{\rm
eff}$--$\log g$--color scales based on synthetic colors calculated
with different stellar atmosphere codes is very good, with typical
differences being well within $\Delta T_{\rm eff} \sim 70$\,K at
$\MoH=-1.0$ and $\Delta T_{\rm eff} \sim 40$\,K at $\MoH=-2.0$.

}

\maketitle

\vskip6cm

\section{Introduction}

Stars on the red giant branch and asymptotic giant branch (RGB and
AGB, respectively) are important constituents of intermediate age
and old stellar populations. In this age range they contribute
significantly to the total radiative energy output of a given
population, especially at near-infrared wavelengths
\citep[e.g.,][]{ML02}. A realistic representation of the
atmospheres and observed spectral properties of late-type giants
with current stellar atmosphere models is, therefore, of crucial
importance, both for understanding evolution of single stars and
stellar systems. This is especially vital for the studies of
distant stellar populations which have to rely on the most
luminous stars, and frequently on RGB and AGB stars alone.

While theoretical modeling of late-type giant atmospheres has
undergone significant development during the last decade, with
major improvements in the modeling procedure, current stellar
atmosphere models still use a number of simplifications in the
model physics and other assumptions \citep[see, e.g.,][]{G03}.
Indeed, the atmospheres of late-type giants are complex, thus
detailed modeling of certain physical phenomena (convection,
pulsations, shock waves, grain formation, mass loss) should
ideally be done using 3-D radiation hydrodynamics. Obviously,
while the classical 1-D model atmospheres may still be very
valuable in providing the time-averaged properties of late-type
giants (for instance, their broad-band photometric colors), it is
important to know how well these theoretically predicted
quantities reproduce the observations of real stars.

In the first paper of this series \citep[][ Paper~I]{K05} we made
a detailed comparison of synthetic photometric colors produced
using current stellar model atmosphere codes ({\tt PHOENIX}, {\tt
MARCS}, and {\tt ATLAS}) with observations of late-type giants in
the Solar neighborhood, at Solar metallicity. Generally, we found
that observed and theoretical colors agree at the level of
$\pm100$\,K, over a wide range of effective temperatures ($T_{\rm
eff}=3000\dots4800$\,K) and gravities ($\log g=0.0\dots{+3.0}$;
see Paper~I for a detailed discussion). Here we extend this
analysis to sub-Solar metallicities, assuming Solar-scaled
abundances of individual chemical species at $\MoH<0$.

There are 3 parts of this paper. First, we investigate the
influence of metallicity, $\MoH$, on the synthetic broad-band
photometric colors calculated with the {\tt PHOENIX} stellar model
atmosphere code. Second, we derive a set of new $T_{\rm
eff}$--$\log g$ relations at different metallicities, based on the
published spectroscopic effective temperatures and gravities of
152 giants in Galactic globular clusters. We further employ these
relations to derive three new $T_{\rm eff}$--$\log g$--color
scales based on the synthetic colors of {\tt PHOENIX}, {\tt MARCS}
and {\tt ATLAS}, for $\MoH=-0.5$, $-1.0$, $-1.5$, and $-2.0$.
Finally, we provide a detailed comparison of the new $T_{\rm
eff}$--$\log g$--color scales based on the synthetic photometric
colors with a number of $T_{\rm eff}$--color relations available
from the literature. This comparison is done for two
metallicities, $\MoH=-1.0$ and $-2.0$.

The paper is structured as follows. In Sect.~\ref{synthcolors} we
briefly describe synthetic spectra calculated with the {\tt
PHOENIX}, {\tt MARCS} and {\tt ATLAS} model atmospheres, and
outline the procedure used to calculate synthetic photometric
colors. The effects of metallicity on the photometric colors are
discussed in Sect.~\ref{feh}. The new $T_{\rm eff}$--$\log g$
scales for different metallicities are derived in
Sect.~\ref{TGscales}. Here we also discuss a sample of Galactic
globular cluster giants which is employed in the derivation of the
$T_{\rm eff}$--$\log g$ relations. Finally, the new $T_{\rm
eff}$--$\log g$--color relations based on the synthetic colors of
{\tt PHOENIX}, {\tt MARCS}, and {\tt ATLAS} are derived in
Sect.~\ref{teffscales}. A comparison of the new $T_{\rm
eff}$--$\log g$--color scales with $T_{\rm eff}$--color relations
available in the literature is also provided.

\section{Stellar atmosphere models, spectra and synthetic colors of late-type giants\label{synthcolors}}

The comparison of synthetic photometric colors with observations
of late-type giants made in Sect.~\ref{teffscales} utilizes colors
calculated with the {\tt PHOENIX}, {\tt MARCS}, and {\tt ATLAS}
model atmospheres. A detailed description of the model atmosphere
codes can be found in relevant papers ({\tt PHOENIX}: \citet{H03a}
and references therein; {\tt ATLAS}: \citet{CK03}; {\tt MARCS}:
\citet{P03} and \citet{GEE03}). In the following subsections we
briefly summarize only the most crucial issues related to the
calculation of synthetic spectra and broad-band photometric
colors.

\subsection{{\tt PHOENIX} spectra and broad-band colors}

{\tt PHOENIX} photometric colors used in this work were calculated
in Paper~I employing a new grid of {\tt PHOENIX} spectra. This
grid is an update and extension of the previous NextGen library of
synthetic spectra of late-type giants \citep{ng-giants} to lower
effective temperatures and metallicities (Hauschildt et al. 2006,
in preparation\footnote{The spectra
are available at the following URL: ftp://\\
ftp.hs.uni-hamburg.de/pub/outgoing/phoenix/GAIA/v2.6.1/.}). The
atmospheres and spectra in this grid were calculated under the assumption of
spherical symmetry and LTE, for a single stellar mass
$M_{\star}=1\,M_{\odot}$. Microturbulent velocity was set to
$\xi=2$~km\,s$^{-1}$. Typical spectral resolution is $0.2$\,nm in
the optical wavelength range and gradually decreases towards the
infrared wavelengths.

The broad-band photometric colors were calculated in the
Johnson-Cousins-Glass system, using filter definitions from
\citet{B90} for the Johnson-Cousins \emph{BVRI} bands and
\citet{BB88} for Johnson-Glass \emph{JHKL} bands. Conversion of
instrumental magnitudes to the standard Johnson-Cousins-Glass
system was done using zero points derived from the synthetic
colors of Vega (equating all color indices of Vega to zero). The
Vega spectrum used for this purpose was calculated with the {\tt
PHOENIX} code employing full NLTE treatment (see Paper~I for
details).

A detailed description of the {\tt PHOENIX} models, spectra, and
colors is given in Paper~I. The influence of various model
parameters (such as gravity, microturbulent velocity, stellar
mass) on the the broad-band photometric colors is discussed there
as well.

\begin{figure*}[t]
\centering
\includegraphics[width=17.8cm] {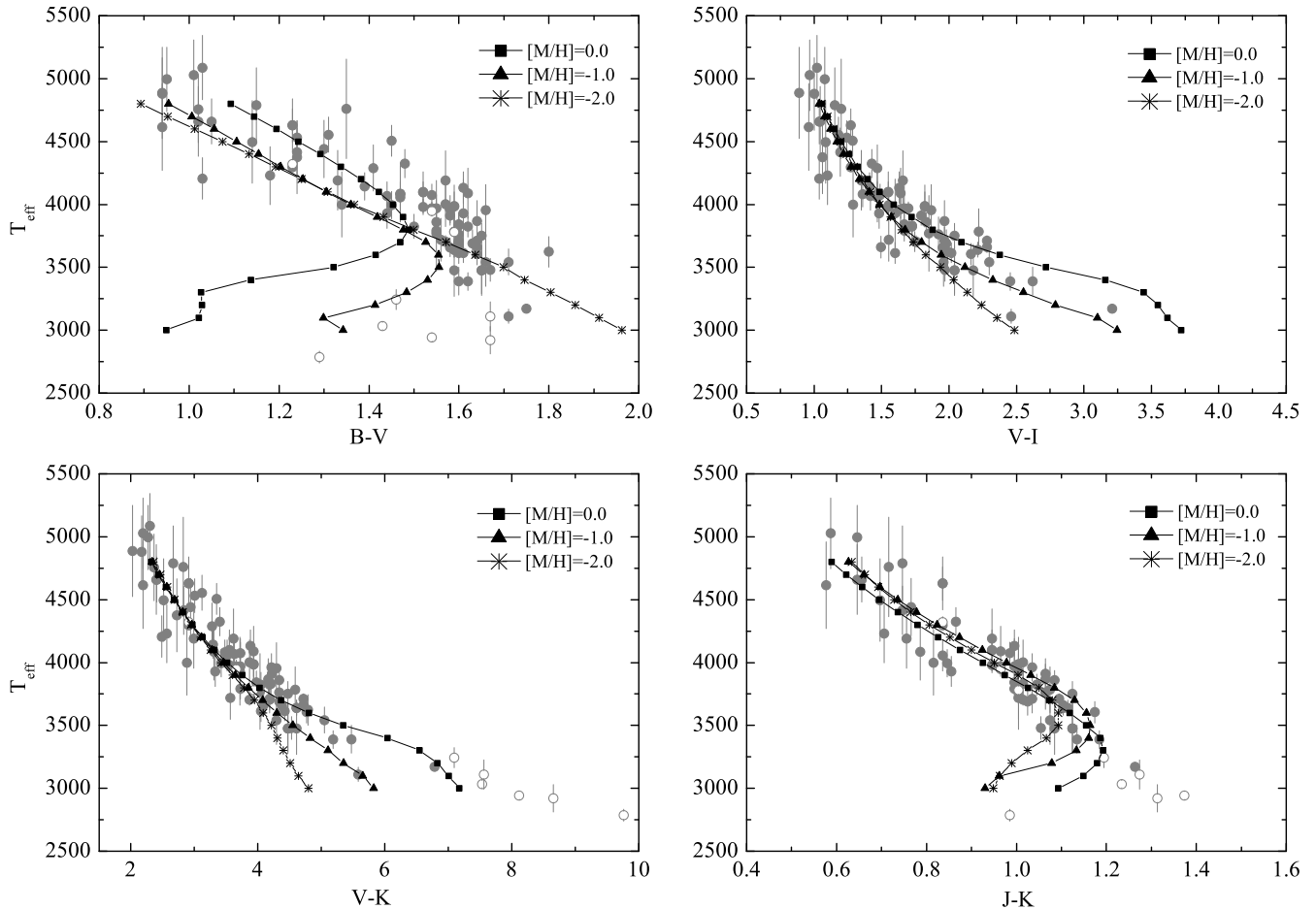}
\caption[]{Influence of metallicity on synthetic photometric
colors in different $T_{\rm eff}$--color planes. Late-type giants
from Paper~I are shown as filled and open circles, indicating
non-variable and variable stars, respectively (see Paper~I for
more details; stars are only plotted to indicate the spread in the
observed $T_{\rm eff}$--color sequences at Solar metallicity, not
for a detailed comparison). Thin lines with symbols are {\tt
PHOENIX} colors at $\log g=1.5$ and different metallicities.
Symbols are spaced at every $100$\,K.\label{fehTC}}
\end{figure*}

\subsection{{\tt MARCS} and {\tt ATLAS} spectra and colors}

Complementary to the new {\tt PHOENIX} colors, we also use
synthetic colors calculated with {\tt MARCS} and {\tt ATLAS}
model atmospheres. {\tt MARCS} spectra were kindly provided to us
by B. Plez (private communication, 2003), {\tt ATLAS} colors were
taken from \citet{CK03}. In both cases synthetic spectra were
calculated in the approximation of plane-parallel geometry, but
employing up-to-date lists of line and molecular opacities
\citep[see e.g.][ for more details]{P03, CK03}.

{\tt MARCS} broad-band photometric colors were calculated in the
Johnson-Cousins-Glass photometric system employing the same
procedure as with {\tt PHOENIX} spectra and using zero points
obtained from the {\tt PHOENIX} spectrum of Vega (Paper~I).

\section{The influence of metallicity on synthetic photometric colors\label{feh}}

It can be anticipated that metallicity has a significant effect on
the synthetic photometric colors of late-type giants, as both
atomic and molecular opacities of various chemical species have a
large influence on the emitted spectrum at the low effective
temperatures typical for late-type giants (Paper~I). In the
following we will focus on the effects due to the variations in
overall metallicity, $\MoH$ (assuming Solar-scaled abundances at
$\MoH<0$); the influence of $\alpha$-element abundances, $\aoFe$,
will be discussed in a separate paper (Ku\v{c}inskas et al. 2006,
in preparation).

\begin{figure}[tbp]
\centering
\includegraphics[width=8.8cm] {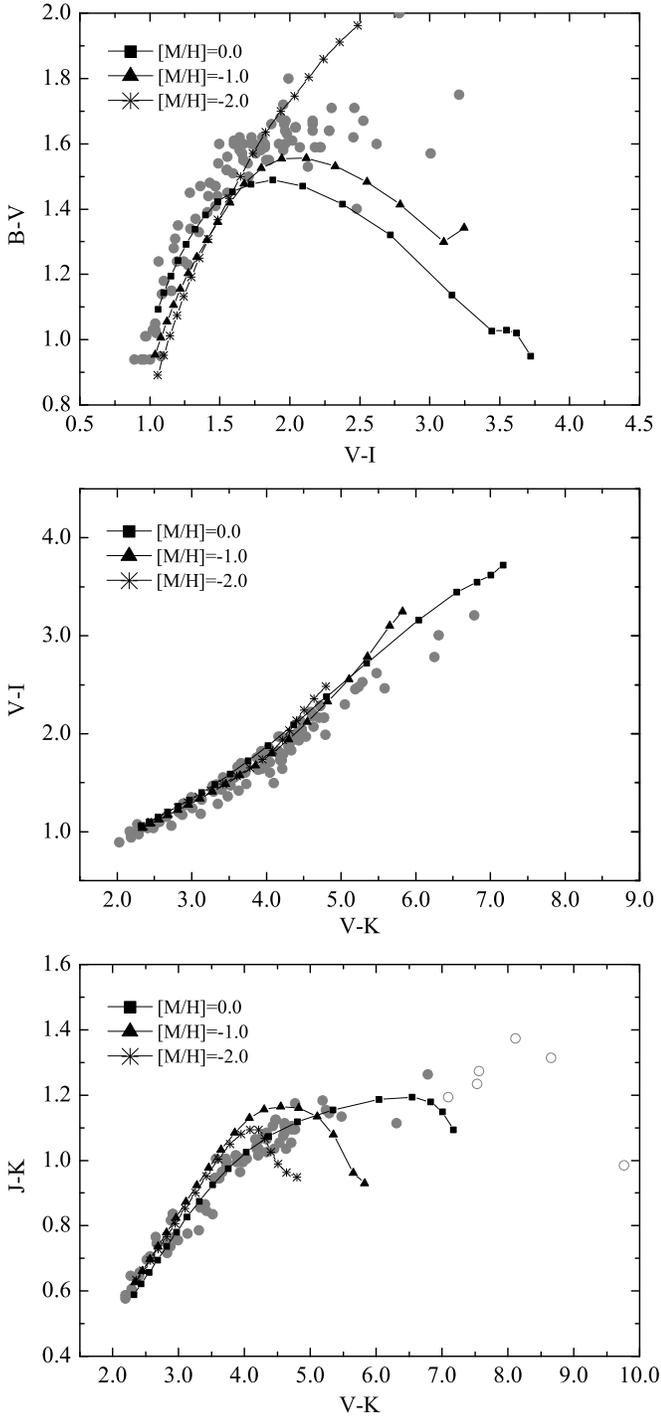}
\caption[]{Same as Fig.~\ref{fehTC} but in the color--color
planes.\label{fehCC}}
\end{figure}

The influence of metallicity on the broad-band photometric colors
is shown in Fig.~\ref{fehTC} ($T_{\rm eff}$--color planes) and
Fig.~\ref{fehCC} (color--color planes). Together with synthetic
colors of {\tt PHOENIX} at several metallicities ($\MoH=0, -1.0,
-2.0$; $\log g=1.5$), the figures also display observations of
individual late-type giants at Solar metallicity from Paper~I
(with $T_{\rm eff}$ available from interferometry), to illustrate
the typical scatter in the observed sequences in $T_{\rm
eff}$--color and color--color planes.

Generally, the effects of metallicity are small in all colors of
the $T_{\rm eff}$--color planes for $T_{\rm eff}\ga4000$\,K. Most
insensitive to the changes in $\MoH$ is $V-K$, which remains
essentially unaffected at $T_{\rm eff} \ga 3700$\,K. This behavior
simply reflects the fact that photometric colors are little
affected by molecular opacities at higher effective temperatures
(Paper~I), thus the trends in the $T_{\rm eff}$--color planes are
governed by changes in atomic opacities which have relatively
little influence on the broad-band photometric colors (note that
these effects become non-negligible at shorter wavelengths,
$\lambda\la450$\,nm).

However, the importance of molecular opacities, and thus the
sensitivity to metallicity effects, increases rapidly at lower
effective temperatures. This is especially pronounced in the
$T_{\rm eff}$--$(B-V)$ plane, where photometric colors develop a
strong dependence on metallicity below $T_{\rm eff}\sim3800$\,K
(with redder $B-V$ colors at lower $\MoH$ values in this $T_{\rm
eff}$ range; note that the trend is opposite at higher
temperatures). The `turn-off' towards the bluer colors (which sets
in at $T_{\rm eff}\sim3800$\,K at $\MoH=0.0$) also tends to occur
at lower $T_{\rm eff}$ and redder colors at lower metallicities.
As the `turn-off' in the $T_{\rm eff}$--$(B-V)$ plane is caused by
the rapidly increasing TiO absorbtion in the $V-$band with
decreasing $T_{\rm eff}$ (see Paper~I), an increasingly larger
fraction of the available Ti is bound in TiO to produce the same
band strength at lower $\MoH$. This causes the `turn-off' to shift
to lower effective temperatures with decreasing metallicity, as
the efficiency of TiO formation (and thus the concentration of TiO
molecules) grows rapidly with decreasing $T_{\rm eff}$.

An increasing influence of metallicity on broad-band photometric
colors at lower effective temperatures can be seen in other
$T_{\rm eff}$--color planes too, with colors becoming bluer at
lower $\MoH$. Typically, this is due to the decreasing
concentration of various molecules at a given effective
temperature with decreasing $\MoH$. For instance, the blueward
shift in the $T_{\rm eff}$--$(V-I)$ and $T_{\rm eff}$--$(V-K)$
planes at $T_{\rm eff}\la 4000$\,K is essentially governed by the
decreasing abundance of TiO at lower $\MoH$ values (at $T_{\rm
eff} \la 3700$\,K the effect of lower VO and ${\rm H}_{2}{\rm O}$
concentrations becomes important in $I$ and $K$ bands,
respectively). The trends seen in the $T_{\rm eff}$--$(J-K)$ plane
are caused by a complex interplay of decreasing concentrations of
${\rm H}_{2}{\rm O}$, CO and TiO (see Paper~I for a discussion on
the influence of molecular opacities on the photometric colors).

The trends seen in the color--color diagrams essentially reflect
the behaviour in the corresponding $T_{\rm eff}$--color planes.
The influence of metallicity is strong in the $(B-V)-(V-I)$ plane
for $V-I \ga 1.7$; a similar effect is seen in the $(J-K)-(V-K)$
plane for $V-K \ga 4.0$. Note that the effects of metallicity are
minor in the $(V-I)-(V-K)$ plane. Interestingly, this color--color
plane is also little affected by gravity and the choice of
microturbulent velocity (Paper~I).

It should be noted that the minimum $T_{\rm eff}$ for which
metallicity effects are negligible tends to increase with gravity.
For instance, at $\log g=0.5$, the effects of $\MoH$ in $V-K$ are
indeed minor for $T_{\rm eff} \ga 3700$\,K, while at $\log g=2.5$
they are only negligible for $T_{\rm eff} \ga 4000$\,K. This is
due to the fact that at temperatures close to where the molecules
dissociate, the effect of metallicity is rather large, as the
pressure in the line forming regions increases with lower $\MoH$
due to the overall lower opacities. This causes a shift in the
chemical equilibria in the line forming regions, which is more
pronounced at higher gravities (where the relative changes in the
pressure are higher) than at low gravities where less molecules
form (for given temperature and metallicity).

\begin{table*}[tb]
\caption[]{Properties of the Galactic globular cluster sample used
in the derivation of $T_{\rm eff}$--$\log g$ relations (see text
for details).\label{GGCsample}}

\centering
\begin{tabular}{clcccccll}
\hline \noalign{\smallskip}
 {\rm Group} & {\rm Cluster} & $n_{star}$ & $n_{spec}$ & $N$ & $\langle\FeoH_{\rm CG97}\rangle$ & $\langle\FeoH_{\rm other}\rangle$ & Reference & Age, Gyr \\
\noalign{\smallskip} \hline \noalign{\smallskip}
  1  & M71, NGC 104       &  21  &  10  &  36  &  $-0.74 \pm 0.12$  &  $-0.81 \pm 0.08$  &  M71: CG97, KI03, R01  & M71:      $10.2\pm1.4$  \\
     &                    &      &      &      &                    &                    &  NGC 104: CG97         & NGC 104:  $10.7\pm1.0$  \\
  2  & M4, M5             &  31  &  17  &  31  &  $-1.11 \pm 0.12$  &  $-1.17 \pm 0.03$  &  M4: I99               & M4:       $11.7\pm0.8$  \\
     &                    &      &      &      &                    &                    &  M5: CG97              & M5:       $11.3\pm1.1$  \\
  3  & M10, M13,          &  46  &  35  &  70  &  $-1.40 \pm 0.09$  &  $-1.58 \pm 0.09$  &  M10: KI03             & M10:      $12.0\pm1.1$  \\
     & NGC 7006           &      &      &      &                    &                    &  M13: CG97, KI03       & M13:      $12.5\pm1.2$  \\
     &                    &      &      &      &                    &                    &  NGC 7006: KI03        & NGC 7006: $13.1\pm1.0$  \\
  4  & NGC 6397           &  10  &  5   &  18  &  $-1.81 \pm 0.13$  &  $-1.98 \pm 0.02$  &  CG97, MPG96           & NGC 6397: $12.3\pm1.1$  \\
  5  & M15, M92           &  54  &  15  &  66  &  $-2.16 \pm 0.06$  &  $-2.36 \pm 0.05$  &  M15: KI03, S00        & M15:      $11.8\pm0.8$  \\
     &                    &      &      &      &                    &                    &  M92: CG97, KI03, S00  & M92:      $12.6\pm0.9$  \\
\noalign{\smallskip} \hline
\end{tabular}

\end{table*}

\section{New $T_{\rm eff}$--$\log g$ scales at $\MoH<0$\label{TGscales}}

A detailed comparison of synthetic photometric colors with
observations in the $T_{\rm eff}$--color planes can be made in a
straightforward way if synthetic colors are provided in the form
of $T_{\rm eff}$--$\log g$--color relations, with gravities at
different temperatures specified according to a representative
$T_{\rm eff}$--$\log g$ relation. Such an approach was adopted in
Paper~I to make a comparison of synthetic and observed colors at
Solar metallicity. In this work we employ the same strategy to
compare photometric colors of late-type giants at sub-Solar
metallicities.

Since a homogeneous set of $T_{\rm eff}$--$\log g$ relations
suitable for use with late-type giants at sub-Solar metallicities
is not readily available in the literature, we focus in this
Section on the derivation of new empirical $T_{\rm eff}$--$\log g$
scales, at $\MoH=-0.5, -1.0, -1.5, -2.0$. For this purpose we
employ a sample of late-type giants in Galactic globular clusters
(GGCs), with their effective temperatures and gravities collected
from the literature. The choice of GGC giants is based on several
considerations. First, GGCs are generally well observed, with
precise atmospheric parameters of individual stars available over
a wide range of effective temperatures and gravities. Second, GGCs
span a wide metallicity range, which is crucial for the derivation
of $T_{\rm eff}$--$\log g$ relations at different $\MoH$. Finally,
they are of similar age which implies a similarity in the
atmospheric parameters of individual stars in clusters of similar
metallicity. This allows binning of stars from different clusters
into groups according to their metallicities, to increase the
number of individual stars available in a given metallicity group.

\subsection{Stellar sample \label{TGsample}}

Ideally, the new $T_{\rm eff}$--$\log g$ relations should be
derived using a sample of late-type giants with their effective
temperatures and gravities measured using direct methods, e.g.,
effective temperatures obtained via the interferometric
measurements of stellar radii, and gravities from parallaxes
(i.e., `evolutionary' gravities, see Sect.~\ref{GGCloggs} below).
Unfortunately, direct measurements of atmospheric parameters are
currently available only for the nearby giants, the majority of
which have metallicities close to Solar. Needless to say,
late-type giants in the GGCs are not accessible, due to the large
distances involved. For this reason, our new relations derived in
this Section are based on spectroscopic effective temperatures and
gravities of GGC giants (in combination with $T_{\rm eff}$ and
$\log g$ from photometry, see below), with temperatures obtained
under the assumption of excitation equilibrium of Fe~I lines, and
gravities derived assuming the ionization equilibrium of Fe~II
lines.

The core of our sample is based on the GGCs analyzed by the
Lick-Texas group, in particular, their recent spectroscopic study
of late-type giants in 16 GGCs \citep[][ KI03, and references
therein]{KI03}. Whenever available, data from other sources were
also used, providing a consistency check between the results of
different groups. We made every attempt to assure that the $T_{\rm
eff}$ used in our study are derived from the analysis of recently
obtained high quality spectroscopic data, or literature data
re-analyzed in a self-consistent and homogeneous way, employing
the advantages of improved stellar model atmospheres, analysis
techniques, and so forth.

The final sample consists of 82 GGC late-type giants with precise
estimates of effective temperatures, gravities and metallicities
from high-resolution spectroscopy. Additionally, we use 70 stars
with $T_{\rm eff}$ and $\log g$ available from photometry, which
nearly doubles the number of stars and individual measurements of
stellar parameters available for the analysis. We justify the
latter choice by making a careful verification of the effective
temperatures and gravities derived with photometry, spectroscopy
and the infrared flux method (IRFM; Blackwell et al. 1990), finding a
generally good agreement in $T_{\rm eff}$ and $\log g$ obtained
with these different methods (Sect.~\ref{GGCteffs} and
Sect.~\ref{GGCloggs}). Metallicities of the individual sample
stars are in the range of $\FeoH=-0.7\dots-2.5$.

Individual clusters with similar $\FeoH$ values were binned into
five metallicity groups, to increase the number of stars in each
group. In fact, this procedure may smear intrinsic morphological
differences in the giant branches of different clusters in the
$T_{\rm eff}$--$\log g$ plane. However, as will be shown in
Sect.~\ref{teffscales}, differences between the RGB sequences of
individual clusters in the $T_{\rm eff}$--$\log g$ plane are
always considerably smaller than the typical errors in the
derivations of $T_{\rm eff}$ and $\log g$.

Properties of the five cluster groups are summarized in
Table~\ref{GGCsample}. The contents of columns 3--5 are: $n_{\rm
star}$ is a total number of stars in a given group for which a
total of $N$ (spectroscopic plus photometric) independent
derivations of $T_{\rm eff}$ and $\log g$ is available; $n_{\rm
spec}$ gives the number of stars with both effective temperatures
and gravities available from high-resolution spectroscopy.

Below we focus on the atmospheric parameters of the sample stars
in more detail.

\subsubsection{Effective temperatures\label{GGCteffs}}

Approximately half of the individual cluster giants used in
this study (77 stars) have effective temperatures available from
spectroscopic analysis (i.e., derived under the assumption of
excitation equilibrium of Fe~I lines). However, spectroscopic
estimates are rather sensitive to different model assumptions,
details of the analysis procedure, etc. For example, \citet{TI99}
have shown that at low metallicities Fe~I lines may suffer
considerably from overionization, due to the leakage of UV photons
into the outer stellar atmosphere because of the lower opacities
associated with lower $\FeoH$ values. This may indeed have an
effect if the effective temperatures are derived under the
assumption of excitation equilibrium of Fe~I.

\begin{figure}[tb]
\includegraphics[width=8.8cm] {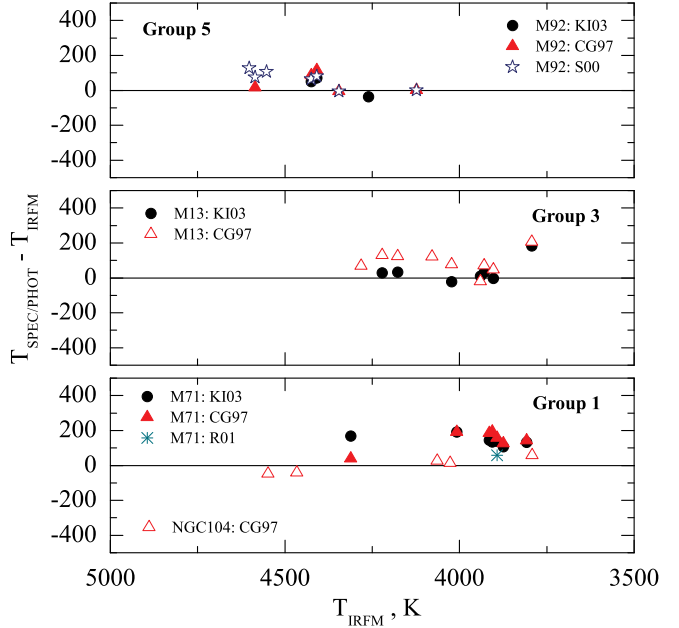}
\caption[]{Comparison of the IRFM temperatures of late-type giants
in Galactic globular clusters used in this work, as derived by
\citet[][ A99]{A99}, with $T_{\rm eff}$ obtained from spectroscopy
and photometry (cluster groups 5, 3, 1, top-down; the contents and
properties of individual cluster groups are given in
Table~\ref{GGCsample}). Different symbols indicate original
sources from which spectroscopic (filled circles) and photometric
(other symbols) effective temperatures were taken. Note that in
case of groups 5 and 3 IRFM temperatures are only available for
stars in one cluster. \label{figTeffs}}
\end{figure}

For seventeen late-type giants in three clusters from
Table~\ref{GGCsample} (M92, M13, M71), effective temperatures are
available both from spectroscopy (KI03) and the infrared flux
method, IRFM \citep[][ A99]{A99}. Effective temperatures and
angular diameters obtained with the IRFM usually agree fairly well
with direct estimates available from interferometry and/or lunar
occultations, for stars of different luminosity classes and
effective temperatures \citep[see,
e.g.,][]{N01,BLG98,AAM99,ASA00}. A comparison with the IRFM
temperatures may thus offer an independent sanity check for the
spectroscopically and photometrically derived effective
temperatures. This is done in Fig.~\ref{figTeffs}.

Clearly, the differences between $T_{\rm eff}$ from spectroscopy
and $T_{\rm IRFM}$ (filled circles in Fig.~\ref{figTeffs}) are
small in the cases of M92 and M13; the average offsets are $\Delta
T_{\rm eff}\simeq30$\,K and $\Delta T_{\rm eff}\simeq40$\,K,
respectively, (i.e., spectroscopic $T_{\rm eff}$ are higher), with
RMS residuals of $\simeq60$\,K and $\simeq70$\,K. There is no
evidence for any statistically significant systematic trends
either. Note, however, that in the case of M13 the average offset
is mostly determined by the large offset of a single star at
$T_{\rm IRFM}=3790\,K$ ($\Delta T_{\rm eff}\simeq180$\,K), and
amounts to only $\Delta T_{\rm eff}\simeq10$\,K if this star is
removed from the averaging procedure. A larger offset is seen in
case of M71, with spectroscopic temperatures derived by KI03 being
higher by $\Delta T_{\rm eff}\simeq140$\,K (RMS residual
$\simeq30$\,K).

A comparison of photometrically derived effective temperatures of
stars in our sample (available for 33 objects) with those obtained
with the IRFM shows that photometric $T_{\rm eff}$ generally tend
to be slightly higher, with the average offsets of about $85$\,K,
$90$\,K and $55$\,K for clusters in groups 5, 3, and 1,
respectively (triangles and asterisks in Fig.~\ref{figTeffs}).
While the average offset in group 1 is relatively small, the RMS
residual is large ($\simeq150$\,K), due to the large offset of the
photometric temperatures of individual stars in M71
($\simeq150$\,K) derived by \citet[][ CG97]{CG97}.

Interestingly, while KI03 and CG97 stars in M71 are obvious
outliers in Fig.~\ref{figTeffs}, they all lie very close to the
average $T_{\rm eff}$--$\log g$ sequence in the $T_{\rm
eff}$--$\log g$ plane (Sect.~\ref{TGrelations},
Fig.~\ref{figTGscales}). At the same time, there are no signs for
the discrepancies between the spectroscopic and evolutionary
gravities of these stars (Sect.~\ref{GGCloggs}). Also, their
gravities agree well with those of stars from other clusters in
this metallicity group. Altogether this may indicate that
discrepancies seen in Fig.~\ref{figTeffs} may in fact be due to
somewhat lower IRFM temperatures of A99 rather than the
systematically higher temperatures of KI03 and CG97.

One should note, however, that the number of stars used in this
comparison is small, thus the trends hinted at in
Fig.~\ref{figTeffs} may clearly be influenced (or even governed)
by selection effects. KI03, for instance, find a good agreement
between spectroscopically obtained effective temperatures and
those resulting from photometry for the majority of late-type
giants in their study (containing 149 stars). Nevertheless, the
differences in $T_{\rm eff}$ of individual stars derived by
different authors may provide an idea about the limits of the
precision of photometrically derived effective temperatures,
reflecting the influence of various systematical effects,
discrepancies in $T_{\rm eff}$ predicted by different $T_{\rm
eff}$--color relations, and so forth. It is obvious that in many
cases these differences are considerably larger than $\sim100$\,K,
a value frequently quoted as a typical uncertainty for the
photometrically derived effective temperatures, which suggests
that errors in the photometrically derived $T_{\rm eff}$ are
frequently seriously underestimated (see also discussion in
Paper~I).

\subsubsection{Gravities\label{GGCloggs}}

Two kinds of gravities are used in spectroscopic abundance
analyses. One is `evolutionary' gravity, which is derived through
the $T_{\rm eff}$--$L_{\star}$--$M_{\star}$--$\log g$ relation
($L_{\star}$ and $M_{\star}$ are stellar luminosity and mass,
respectively), where effective temperature is typically obtained
from photometric colors, luminosity from absolute magnitude and
bolometric correction, and the mass is implied from evolutionary
tracks. The major disadvantage of evolutionary gravities from the
point of view of our study is that an estimate of gravity through
the $T_{\rm eff}$--$L_{\star}$--$M_{\star}$--$\log g$ relation
makes an implicit use of the $T_{\rm eff}$--$\log g$ scale for the
derivation of stellar mass. Fortunately, since all clusters in our
sample are old (see Sect.~\ref{GGCages}), masses of individual
stars on their giant branches should be very similar. In such a
situation, the $T_{\rm eff}$--$L_{\star}$--$M_{\star}$--$\log g$
relation becomes essentially independent of stellar mass.
Evolutionary models predict that for clusters with ages of $10$
and $13$\,Gyr (the limits for the clusters in our sample,
Sect.~\ref{GGCages}) the difference in $\log g$ because of the
difference in mass of the RGB stars ($\la0.1$\,$M_\odot$) does not
exceed $\sim0.04$\,dex \citep{YY01}. Thus, the gravity estimate
will basically rely only on the estimate of effective temperature
and stellar luminosity.

Another type of gravity estimate is obtained directly from the
spectral analysis, i.e., derived under the assumption of
ionization equilibrium of Fe I and Fe II lines. These are referred
to as `spectroscopic' gravities, and in this case no assumptions
regarding the $T_{\rm eff}$--$\log g$ scale are made. However, the
assumption of ionization equilibrium for Fe I and Fe II in the
atmospheres of late-type stars might be a rather poor
approximation. As was mentioned in Sect.~\ref{GGCteffs}, Fe I
lines (especially in metal-poor stars) may form in conditions that
are far from the LTE, due to overionization of Fe I by ultraviolet
radiation \citep{TI99}. As a consequence, surface gravities
derived under this assumption may also be in error. \citet{NHS97}
have shown that there indeed exists a systematic discrepancy
between spectroscopic gravities and those obtained using {\it
HIPPARCOS} parallaxes. Unfortunately, the samples discussed by
\citet{NHS97} do not extend into the gravity range of late-type
giants. Only three stars have $\log g<3.0$ in the sample of
\citet{TI99}, however, the differences between LTE and NLTE
gravities for these stars are less than $0.1$\,dex.

\begin{figure}[tb]
\includegraphics[width=8.8cm] {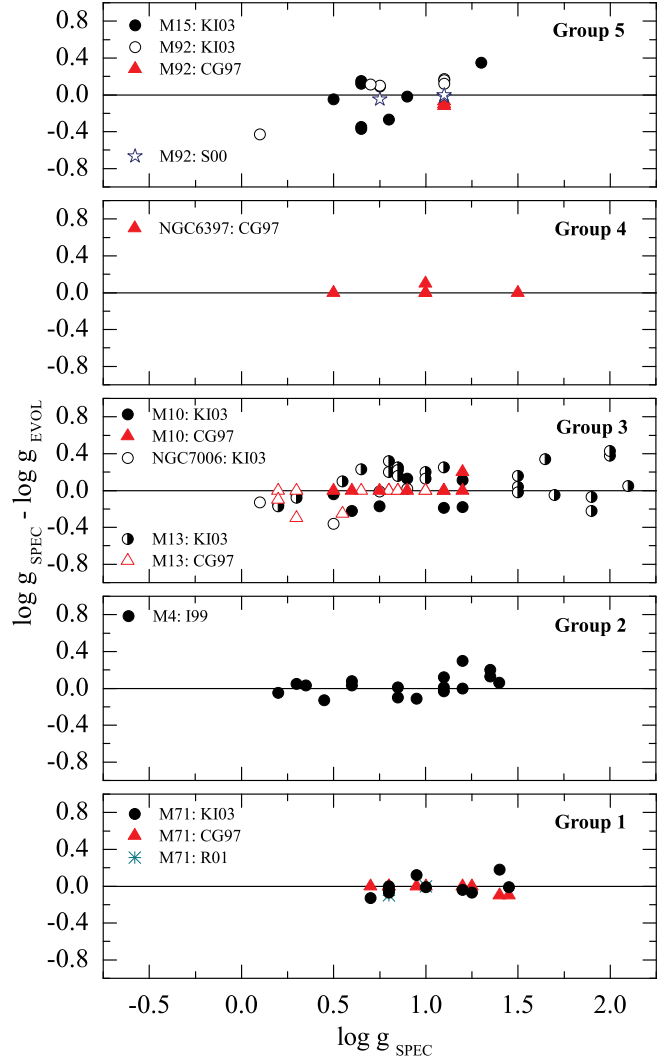}
\caption[]{The difference between spectroscopic and evolutionary
gravities of Galactic globular cluster giants used in this work,
for clusters in groups 5 to 1 in Table~\ref{GGCsample} (top-down).
Spectroscopic gravities are from MPG96 for NGC6397, I99 for M4,
and from KI03 for the rest of cluster giants. Different symbols
indicate original sources from which evolutionary gravities were
taken (see text for details). Note that in case of groups 4, 2, 1
spectroscopic and evolutionary gravities are available for stars
in only one cluster. \label{figGGCloggs}}
\end{figure}

A detailed comparison of spectroscopic and evolutionary gravities
was recently made by KI03 for a large sample of late-type giants
in 16 GGCs. They find a good agreement between the two types of
gravity estimates within a large range of gravities, effective
temperatures and metallicities. Similar conclusions were reached
also by \citet[][ I99]{I99} and \citet{I01} in their studies of
late-type giants in the globular clusters M4 and M5, respectively.

The conclusions of KI03 and I99 are well reflected in
Fig.~\ref{figGGCloggs}, where we compare spectroscopic and
evolutionary gravities of stars in the GGCs listed in
Table~\ref{GGCsample}. The differences between spectroscopic and
evolutionary gravities are calculated using spectroscopic $\log g$
from \citet[][ MPG96]{MPG96} for NGC6397, I99 for M4, and KI03 for
the rest of cluster giants. Evolutionary gravities were taken from
a number of sources (CG97; I99; KI03; Ramirez et al. 2001, R01;
Sneden et al. 2000, S00). The agreement between spectroscopic and
evolutionary gravities is very good within a large interval of
gravities and metallicities, $\log g \sim 0.2\dots2.2$ and
$\FeoH\sim-0.7\dots-2.5$, with no indication of statistically
significant offsets or systematical trends
(Fig.~\ref{figGGCloggs}). There are no obvious trends between the
data of different authors either. Formal linear fits drawn through
the data in Fig.~\ref{figGGCloggs} do not deviate from $(\log
g)_{\rm SPEC} - (\log g)_{\rm EVOL}=0$ by more than $0.2$\,dex
within the entire interval of gravities and metallicities. The RMS
scatter of the data around $\Delta \log g=0$ is largest for stars
in cluster group 3 (M10, M13, NGC 7006), $\simeq0.2$\,dex, which
is similar to a typical error margin of spectroscopically derived
gravities ($\sim0.2$\,dex).

In case of several clusters (M15, M92, M13, NGC 7006) there is an
indication that at $\log g\la0.8$ evolutionary gravities of some
stars may be somewhat higher than those obtained from
spectroscopy. One possible explanation for this slight discrepancy
may be that some of these stars are in fact on the AGB, not the
RGB. Since the outer atmospheres may be more extended in case of
AGB stars, spectroscopic gravities may indicate lower effective
$\log g$ than what would be inferred from the $T_{\rm
eff}$--$L_{\star}$--$M_{\star}$--$\log g$ relation.

\subsubsection{Metallicities and ages\label{GGCages}}

\begin{figure}[b]
\includegraphics[width=8.8cm] {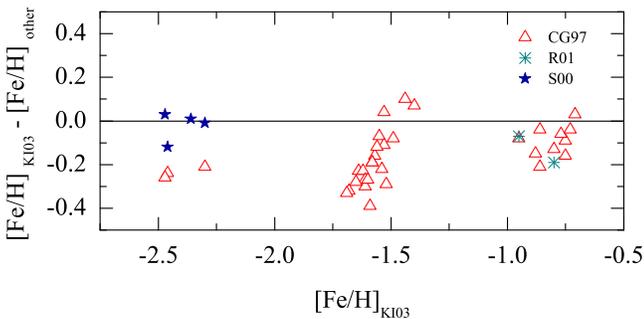}
\caption[]{The difference between metallicities derived by KI03
and those obtained in other studies for the cluster giants used in
this work, plotted against $\FeoH$ of KI03. Different symbols
indicate the original sources from which the $\FeoH$ values were
taken.\label{FoHdiffs}}
\end{figure}

Metallicities of individual stars in groups 1, 3, 5
(Table~\ref{GGCsample}) are compared in Fig.~\ref{FoHdiffs}, which
shows the difference between $\FeoH$ estimates obtained by KI03
and those derived by other authors. Whenever available, we used
metallicities derived from the Fe~II lines, to avoid possible bias
due to inadequate treatment of Fe~I lines with LTE model
atmospheres of late-type giants (cf. Thevenin \& Idiart 1999; see
also KI03 for a discussion).

There is a clear indication that Fe abundances derived by CG97 are
systematically higher than those obtained by KI03, by
$\simeq0.16$\,dex on average (with an RMS scatter of
$\pm0.12$\,dex). Similarly, $\FeoH$ values of CG97 are higher than
those derived by MPG96 by $\simeq0.11$\,dex. On the other hand,
the data from MPG96 and R01 are in good agreement with the $\FeoH$
estimates of KI03. Since the abundances in KI03 seem to agree well
with the $\FeoH$ scales of \citet{ZW84}, and \citet{R97}, it is
most likely that these differences simply reflect a well known
discrepancy between the metallicity scales of \citet{ZW84} and
CG97. We, therefore, derive two average $\FeoH$ values for each
cluster group: one estimate based on CG97 metallicities, the other
on KI03, combined with data from other sources (MPG96, S00, R01).
Only in groups 1 and 2 is there a good agreement between the two
scales; differences within the other groups are marginally
significant (at the $1-2\sigma$ level). It should be mentioned
though, that in all cases the spread in metallicities within a
particular cluster group is small; typically, $\sim95\%$ of stars
are within a $\pm 0.2$\,dex margin from the mean values given in
Table~\ref{GGCsample}.

Ages of the individual clusters  (taken from Salaris \& Weiss
2002) are provided in Table~\ref{GGCsample} (age estimate for
NGC~7006 is from Santos \& Piatti 2004). The cluster ages of
\citet{SW02} were derived from the difference in luminosity of the
horizontal branch and the main sequence turn-off point in the
cluster color-magnitude diagram, using both CG97 and \citet{ZW84}
metallicity scales. The differences between ages corresponding to
the two metallicity scales are small: for all clusters in
Table~\ref{GGCsample} they are well within $\sim0.5$\,Gyr
\citep{SW02}, thus averaged values are given in
Table~\ref{GGCsample}.

\begin{figure*}[t]
\sidecaption
\includegraphics[width=12.1cm] {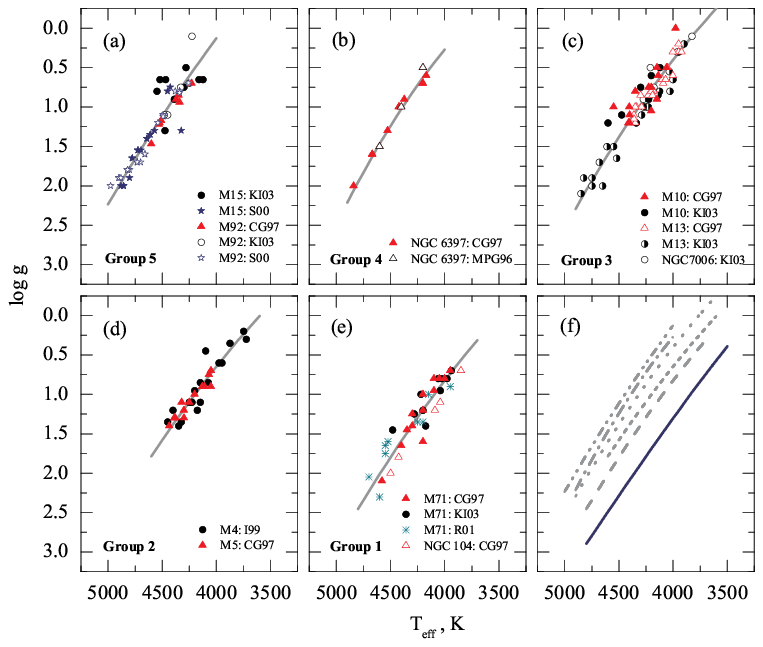}
\caption[]{~Empirical $T_{\rm eff}$--$\log g$ relations (solid
lines) obtained as best fits to the observed late-type giant
sequences in the five cluster groups (Table~\ref{GGCsample}).
Panels (a)--(e) show data for groups 5 to 1 respectively; panel
(f) displays all derived $T_{\rm eff}$--$\log g$ relations (groups
5 to 1, top-down), together with the $T_{\rm eff}$--$\log g$ scale
of \citet[][ H00]{H00} for $\FeoH=0.0$ (thick solid line).
Different symbols indicate original sources from which the data
were taken.\label{figTGscales}}
\end{figure*}

All clusters in our sample are old, with individual ages between
$\sim10-13$\,Gyr (Table~\ref{GGCsample}). The shift between the
RGB isochrones corresponding to these limiting ages is $\Delta
T_{\rm eff}<100$\,K at $\FeoH=-0.7$, and decreases with lower
$\FeoH$ (Yi et al. 2001). While the age differences may indeed
introduce additional scatter in the $T_{\rm eff}$--$\log g$ plane,
no clear indication for such spread is seen in the observed
sequences of different cluster groups (Fig.~\ref{figTGscales},
Sect.~\ref{TGrelations}), most likely because these differences
are smeared out by the larger errors in spectroscopically or
photometrically derived effective temperatures and/or gravities.

\subsection{New $T_{\rm eff}$--$\log g$ scales\label{TGrelations}}

The gravities of individual stars in all five cluster groups
(Table~\ref{GGCsample}) are plotted versus the effective
temperature in Fig.~\ref{figTGscales}. Generally, there is good
consistency in $T_{\rm eff}$ and $\log g$ of individual giants
within a particular cluster group (even though these stars belong
to different clusters), especially in groups 2, 4, and 5. Several
stars at $T_{\rm eff} \sim 4500$\,K in M15 (group 5,
Fig.~\ref{figTGscales}a) are somewhat off the main trend, as their
spectroscopic gravities are considerably lower than those of other
stars in this effective temperature range; similarly deviating is
one star in M92 at $T_{\rm eff} \sim 4200$\,K. It should be
reminded that spectroscopic $\log g$ of these stars are about
$0.4$\,dex lower than their evolutionary gravities (see
Fig.~\ref{figGGCloggs}). Note however, that the resulting $T_{\rm
eff}$--$\log g$ scale for this cluster group remains essentially
unaffected if these stars are not employed in its derivation.
Slightly larger scatter is seen in group 3, though data from
different sources seem to agree well, with no noticeable
differences or trends between them. With the exception of the CG97
data for NGC 104, there is also a good consistency in the
effective temperatures and gravities of individual stars in group
1.

There is a faint hint that the sequence of CG97 stars in M92 is
slightly shifted towards lower effective temperatures with respect
to the best fitting sequence containing all stars in group 5. A
similar offset is more clearly seen for the CG97 data in NGC 104
(group 1), where effective temperatures of CG97 are lower by about
$\sim 150$\,K. Fortunately, in both cases these offsets have minor
impact on the resulting $T_{\rm eff}$--$\log g$ scales (which are
not altered significantly if these stars are excluded), and we
therefore retain them in the further analysis.

The new empirical $T_{\rm eff}$--$\log g$ relations obtained as
best fits to the observed giant sequences in the five cluster
groups (Table~\ref{GGCsample}) are shown in Fig.~\ref{figTGscales}
and are provided in numerical form in Tables~\ref{empGGCscales}
and \ref{TGscalefits} (Table~\ref{TGscalefits} lists coefficients
of polynomial fits representing the new relations). The typical
RMS residual of the best fit is $\simeq0.15$\,dex in $\log g$ (see
Table~\ref{empGGCscales}), which corresponds to about
$\simeq100$\,K in $T_{\rm eff}$ (note that the typical
uncertainties in spectroscopically and/or photometrically derived
$T_{\rm eff}$ and $\log g$ are somewhat larger, $\sim150$\,K and
$\sim0.3$\,dex, respectively). This uncertainty (RMS residual) is
slightly smaller than the typical errors in the $T_{\rm
eff}$--color relations of A99 that are based on stellar
temperatures derived with the IRFM (up to $\sim150$\,K), and the
RMS residuals of the $T_{\rm eff}$--color relations based on the
effective temperatures from interferometry ($\sim160$\,K,
Paper~I).

\begin{table}[tb]
\caption[]{Empirical $T_{\rm eff}$--$\log g$ relations for the
late-type giants in different cluster groups, obtained as best
fits to the data in Fig.~\ref{figTGscales}. Column 2 gives $\log
g$ according to the $T_{\rm eff}$--$\log g$ scale of \citet[][
H00]{H00} linearly extrapolated to $T_{\rm eff}=4800$\,K. The last
line is RMS residual of the best-fit, expressed as a gravity
difference. \label{empGGCscales}} \centering
\begin{tabular}{cccccccccccccc}
\hline \noalign{\smallskip}
$T_{\rm eff}$ & \multicolumn{6}{c}{log g}  \\
\noalign{\smallskip} \cline{2-7} \noalign{\smallskip}
        & \FeoH=0.0 & Gr 1 & Gr 2  &  Gr 3  &  Gr 4  & Gr 5   \\
\noalign{\smallskip} \hline \noalign{\smallskip}
  4900  &   --   &   --   &   --   &  2.29  &  2.21  &  2.00  \\
  4800  &  2.90  &  2.46  &   --   &  2.05  &  1.95  &  1.77  \\
  4700  &  2.69  &  2.23  &   --   &  1.82  &  1.70  &  1.55  \\
  4600  &  2.49  &  2.01  &  1.78  &  1.60  &  1.47  &  1.33  \\
  4500  &  2.28  &  1.80  &  1.58  &  1.38  &  1.24  &  1.11  \\
  4400  &  2.08  &  1.59  &  1.38  &  1.17  &  1.02  &  0.91  \\
  4300  &  1.89  &  1.39  &  1.19  &  0.97  &  0.82  &  0.70  \\
  4200  &  1.69  &  1.19  &  1.00  &  0.77  &  0.62  &  0.51  \\
  4100  &  1.50  &  1.01  &  0.82  &  0.58  &  0.44  &  0.31  \\
  4000  &  1.30  &  0.82  &  0.65  &  0.40  &  0.27  &  0.13  \\
  3900  &  1.11  &  0.65  &  0.48  &  0.23  &   --   &   --   \\
  3800  &  0.93  &  0.48  &  0.31  &  0.06  &   --   &   --   \\
  3700  &  0.74  &  0.31  &  0.16  &  -0.10 &   --   &   --   \\
  3600  &  0.57  &   --   &  0.01  &  -0.25 &   --   &   --   \\
  3500  &  0.39  &   --   &   --   &   --   &   --   &   --   \\
  3400  &  0.21  &   --   &   --   &   --   &   --   &   --   \\
  3300  &  0.04  &   --   &   --   &   --   &   --   &   --   \\
\noalign{\smallskip} \hline \noalign{\smallskip}
\multicolumn{2}{l}{RMS residual:}  &  0.18  &  0.12  &  0.15  &  0.06  &  0.16  \\
\noalign{\smallskip} \hline
\end{tabular}
\end{table}

\begin{table*}[tb]
\caption[]{Analytical expressions corresponding to the empirical
$T_{\rm eff}$--$\log g$ relations given in
Table~\ref{empGGCscales}, in the form $\log g = a_0 + a_1 T_{\rm
eff} + a_2 T_{\rm eff}^2 + a_3 T_{\rm eff}^3 + a_4 T_{\rm eff}^4$.
\label{TGscalefits}}

\centering
\begin{tabular}{cccccccll}
\hline \noalign{\smallskip}
 {\rm Group} &    $a_0$    &     $a_1$    &     $a_2$    &     $a_3$     &     $a_4$     & $T_{\rm eff}$ {\rm range}, {\rm K} \\
\noalign{\smallskip} \hline \noalign{\smallskip}
 $\FeoH=0.0$ &     8.07    &  --1.045e-2  &   3.997e-6   &  --5.814e-10  &   3.240e-14   &  $3300 \le T_{\rm eff} \le 4800$ \\
      1      &   --1.41    &  --6.761e-4  &   3.086e-7   &       --      &       --      &  $3700 \le T_{\rm eff} \le 4800$ \\
      2      &   --1.51    &  --6.394e-4  &   2.948e-7   &       --      &       --      &  $3600 \le T_{\rm eff} \le 4600$ \\
      3      &   --0.95    &  --1.100e-3  &   3.597e-7   &       --      &       --      &  $3600 \le T_{\rm eff} \le 4900$ \\
      4      &     2.60    &  --2.820e-3  &   5.596e-7   &       --      &       --      &  $4000 \le T_{\rm eff} \le 4900$ \\
      5      &   --2.87    &  --3.360e-4  &   2.714e-7   &       --      &       --      &  $4000 \le T_{\rm eff} \le 4900$ \\
\noalign{\smallskip} \hline
\end{tabular}

\end{table*}

\begin{table*}[tb]
\caption[]{New $T_{\rm eff}$--$\log g$ relations at several
metallicities, obtained by interpolating empirical $T_{\rm
eff}$--$\log g$ relations from Table~\ref{empGGCscales}. The new
relations are given for the metallicity scales of KI03 and CG97
(see text for details).\label{tabGGCscales}} \centering
\begin{tabular}{ccccccc|ccccc}
\hline \noalign{\smallskip}
 $T_{\rm eff}$  &  \multicolumn{11}{c}{log g}  \\
\noalign{\smallskip} \cline{3-12} \noalign{\smallskip}
    & &  \multicolumn{4}{c}{$\FeoH$ according to}     & & &   \multicolumn{4}{c}{$\FeoH$ according to}     \\
    & &  \multicolumn{4}{c}{CG97 metallicity scale}  & & &   \multicolumn{4}{c}{KI03 metallicity scale}  \\
    & &  --0.5   &   --1.0   &   --1.5   &   --2.0   & & &  --0.5   &   --1.0   &   --1.5   &   --2.0    \\
\noalign{\smallskip} \hline \noalign{\smallskip}
  4900  &  &  2.80  &  2.52  &  2.28  &  2.09  & &  &  2.83  &  2.59  &  2.37  &  2.15   \\
  4800  &  &  2.58  &  2.29  &  2.04  &  1.85  & &  &  2.61  &  2.35  &  2.13  &  1.90   \\
  4700  &  &  2.36  &  2.06  &  1.81  &  1.61  & &  &  2.39  &  2.12  &  1.89  &  1.67   \\
  4600  &  &  2.14  &  1.84  &  1.59  &  1.38  & &  &  2.17  &  1.90  &  1.67  &  1.44   \\
  4500  &  &  1.93  &  1.63  &  1.37  &  1.17  & &  &  1.96  &  1.68  &  1.45  &  1.22   \\
  4400  &  &  1.73  &  1.42  &  1.16  &  0.95  & &  &  1.76  &  1.48  &  1.24  &  1.01   \\
  4300  &  &  1.53  &  1.22  &  0.96  &  0.75  & &  &  1.56  &  1.28  &  1.04  &  0.81   \\
  4200  &  &  1.34  &  1.03  &  0.76  &  0.56  & &  &  1.37  &  1.08  &  0.84  &  0.61   \\
  4100  &  &  1.15  &  0.84  &  0.57  &  0.37  & &  &  1.18  &  0.90  &  0.65  &  0.43   \\
  4000  &  &  0.97  &  0.66  &  0.39  &  0.19  & &  &  1.00  &  0.72  &  0.47  &  0.25   \\
  3900  &  &  0.80  &  0.49  &  0.21  &  0.02  & &  &  0.82  &  0.55  &  0.30  &  0.08   \\
  3800  &  &  0.63  &  0.33  &  0.04  & --0.15 & &  &  0.65  &  0.38  &  0.13  & --0.08  \\
  3700  &  &  0.46  &  0.17  & --0.12 & --0.30 & &  &  0.48  &  0.23  & --0.03 & --0.23  \\
  3600  &  &  0.30  &  0.02  & --0.28 & --0.45 & &  &  0.32  &  0.08  & --0.18 & --0.37  \\
  3500  &  &  0.15  & --0.13 & --0.42 & --0.59 & &  &  0.17  & --0.07 & --0.32 & --0.51  \\
\noalign{\smallskip} \hline
\end{tabular}
\end{table*}

\begin{figure}[tb]
\includegraphics[width=8.8cm] {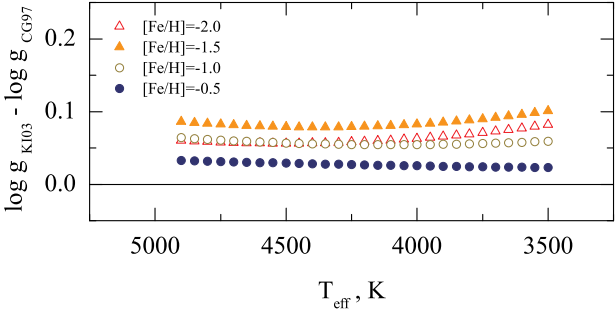}
\caption[]{The difference in gravities predicted by the new
$T_{\rm eff}$--$\log g$ relations (Table~\ref{tabGGCscales}) based
on the metallicity scales of KI03 and CG97, at several
$\FeoH$.\label{difGGCscales}}
\end{figure}

We also provide $T_{\rm eff}$--$\log g$ relations at several
additional metallicities corresponding to the metallicity nodes in
the {\tt PHOENIX}, {\tt MARCS} and {\tt ATLAS} grids of synthetic
photometric colors, at $\MoH=-0.5,-1.0,-1.5,-2.0$. These
supplementary relations were obtained by quadratic interpolation
between the empirical $T_{\rm eff}$--$\log g$ relations in cluster
groups 1-5 (Tables~\ref{empGGCscales} and \ref{TGscalefits}). The
empirical relations were extrapolated to cover the range $T_{\rm
eff}=3500\dots4900$\,K before making the interpolation. Note,
however, that in most cases extrapolation was only done by up to
100\,K in $T_{\rm eff}$, which corresponds to $\sim0.2$\,dex in
$\log g$. Exceptions are relations in groups 4 and 5
($\FeoH<-1.8$), which were extrapolated to lower effective
temperatures by as much as 500\,K, and thus should be used with
caution below $T_{\rm eff}\sim3900$\,K. Metallicities for the
individual cluster groups were assigned according to the
metallicity scales of CG97 and KI03, thus two sets of interpolated
$T_{\rm eff}$--$\log g$ relations corresponding to each
metallicity scale are provided (Table~\ref{tabGGCscales}).
Table~\ref{tabGGCscalesfits} gives analytical expressions
corresponding to the new $T_{\rm eff}$--$\log g$ relations
provided in Table~\ref{tabGGCscales}. It should be noted that
differences between the $T_{\rm eff}$--$\log g$ relations
corresponding to the two metallicity scales are always less than
$0.1$\,dex in $\log g$ (or, correspondingly, $\sim50$\,K in
$T_{\rm eff}$, see Fig.~\ref{difGGCscales}). The RMS residuals of
the interpolation procedure do not exceed $0.03$\,dex in $\log g$.

\begin{table}[tb]
\caption[]{Analytical expressions corresponding to the new $T_{\rm
eff}$--$\log g$ relations given in Table~\ref{tabGGCscales}, in
the form $\log g = a_0 + a_1 T_{\rm eff} + a_2 T_{\rm eff}^2$. \label{tabGGCscalesfits}}

\centering
\begin{tabular}{cccccccll}
\hline \noalign{\smallskip}
      $\FeoH$     &    $a_0$    &     $a_1$     &     $a_2$    \\
\noalign{\smallskip} \hline \noalign{\smallskip}
 {\rm CG97 scale} &             &               &              \\
       -0.5       &   --1.85    &  --3.761e-4   &   2.703e-7   \\
       -1.0       &   --0.82    &  --1.010e-3   &   3.453e-7   \\
       -1.5       &   --1.41    &  --8.991e-4   &   3.371e-7   \\
       -2.0       &   --0.48    &  --1.420e-3   &   3.969e-7   \\
\noalign{\smallskip}
 {\rm KI03 scale} &             &               &              \\
       -0.5       &   --1.81    &  --3.885e-4   &   2.726e-7   \\
       -1.0       &   --0.52    &  --1.130e-3   &   3.600e-7   \\
       -1.5       &   --0.78    &  --1.150e-3   &   3.657e-7   \\
       -2.0       &     0.09    &  --1.650e-3   &   4.225e-7   \\
\noalign{\smallskip} \hline
\end{tabular}

\end{table}

\begin{table*}[t]

\caption[]{$T_{\rm eff}$--$\log g$--color relations for the
late-type giants based on the synthetic colors calculated with
{\tt PHOENIX} and {\tt MARCS} model atmospheres. Photometric
colors are given in the Johnson-Cousins-Glass system (see
Sect.~\ref{synthcolors} for details). \label{TGPhoenixMarcs}}

\centering \scriptsize
\begin{tabular}{ccccccc|cccc}
\hline \noalign{\smallskip}
           &         &     & \multicolumn{4}{c}{\tt PHOENIX} & \multicolumn{4}{c}{\tt MARCS} \\
\cline{4-7} \cline{8-11} \noalign{\smallskip}
 $T_{\rm eff}$ & $\log g$ &  &  $B\!-\!V$ & $V\!-\!I$ & $V\!-\!K$ & $J\!-\!K$ & $B\!-\!V$ & $V\!-\!I$ & $V\!-\!K$ & $J\!-\!K$ \\
\noalign{\smallskip} \hline \noalign{\smallskip}

        &       & &                    \multicolumn{8}{c}{\FeoH=--0.5}                            \\

\noalign{\smallskip}

  4900  &  2.83 & &  0.931  &  0.994  &  2.226  &  0.596  &    --   &    --   &    --   &    --   \\
  4800  &  2.61 & &  0.977  &  1.033  &  2.328  &  0.626  &    --   &    --   &    --   &    --   \\
  4700  &  2.39 & &  1.023  &  1.075  &  2.437  &  0.658  &    --   &    --   &    --   &    --   \\
  4600  &  2.17 & &  1.076  &  1.121  &  2.553  &  0.693  &    --   &    --   &    --   &    --   \\
  4500  &  1.96 & &  1.130  &  1.170  &  2.677  &  0.729  &  1.157  &  1.138  &  2.662  &  0.735  \\
  4400  &  1.76 & &  1.190  &  1.227  &  2.813  &  0.768  &  1.218  &  1.190  &  2.792  &  0.774  \\
  4300  &  1.56 & &  1.250  &  1.288  &  2.958  &  0.810  &  1.279  &  1.245  &  2.932  &  0.816  \\
  4200  &  1.37 & &  1.317  &  1.357  &  3.118  &  0.855  &  1.347  &  1.312  &  3.086  &  0.860  \\
  4100  &  1.17 & &  1.385  &  1.435  &  3.291  &  0.903  &  1.420  &  1.387  &  3.255  &  0.907  \\
  4000  &  1.00 & &  1.452  &  1.523  &  3.479  &  0.953  &  1.492  &  1.472  &  3.439  &  0.957  \\
  3900  &  0.82 & &  1.525  &  1.630  &  3.692  &  1.004  &  1.562  &  1.570  &  3.644  &  1.009  \\
  3800  &  0.65 & &  1.592  &  1.757  &  3.932  &  1.058  &  1.624  &  1.690  &  3.876  &  1.065  \\
  3700  &  0.48 & &  1.650  &  1.915  &  4.212  &  1.114  &  1.672  &  1.841  &  4.148  &  1.121  \\
  3600  &  0.32 & &  1.687  &  2.136  &  4.566  &  1.164  &  1.691  &  2.046  &  4.484  &  1.179  \\
  3500  &  0.17 & &  1.681  &  2.436  &  5.021  &  1.212  &    --   &    --   &    --   &    --   \\

\noalign{\smallskip}

        &        & &                    \multicolumn{8}{c}{\FeoH=--1.0}                            \\

\noalign{\smallskip}

  4900  &  2.59  & &  0.880  &  0.986  &  2.230  &  0.605  &    --   &    --   &    --   &    --   \\
  4800  &  2.35  & &  0.926  &  1.025  &  2.329  &  0.635  &    --   &    --   &    --   &    --   \\
  4700  &  2.12  & &  0.975  &  1.067  &  2.438  &  0.667  &    --   &    --   &    --   &    --   \\
  4600  &  1.90  & &  1.031  &  1.113  &  2.554  &  0.702  &    --   &    --   &    --   &    --   \\
  4500  &  1.68  & &  1.094  &  1.165  &  2.681  &  0.738  &  1.114  &  1.135  &  2.661  &  0.739  \\
  4400  &  1.48  & &  1.157  &  1.220  &  2.816  &  0.778  &  1.183  &  1.185  &  2.783  &  0.774  \\
  4300  &  1.28  & &  1.229  &  1.284  &  2.966  &  0.820  &  1.249  &  1.247  &  2.933  &  0.819  \\
  4200  &  1.08  & &  1.304  &  1.355  &  3.128  &  0.866  &  1.330  &  1.312  &  3.080  &  0.858  \\
  4100  &  0.90  & &  1.384  &  1.433  &  3.302  &  0.914  &  1.402  &  1.387  &  3.255  &  0.910  \\
  4000  &  0.72  & &  1.471  &  1.523  &  3.498  &  0.966  &  1.484  &  1.466  &  3.429  &  0.956  \\
  3900  &  0.55  & &  1.560  &  1.622  &  3.707  &  1.021  &  1.561  &  1.559  &  3.638  &  1.018  \\
  3800  &  0.38  & &  1.659  &  1.744  &  3.945  &  1.078  &  1.641  &  1.663  &  3.851  &  1.071  \\
  3700  &  0.23  & &  1.753  &  1.886  &  4.211  &  1.137  &    --   &    --   &    --   &    --   \\
  3600  &  0.08  & &  1.832  &  2.068  &  4.520  &  1.191  &    --   &    --   &    --   &    --   \\
  3500  & --0.07 & &  1.883  &  2.320  &  4.910  &  1.239  &    --   &    --   &    --   &    --   \\

\noalign{\smallskip}

        &        & &                    \multicolumn{8}{c}{\FeoH=--1.5}                            \\

\noalign{\smallskip}

  4900  &  2.37  & &  0.846  &  0.992  &  2.234  &  0.606  &    --   &    --   &    --   &    --   \\
  4800  &  2.13  & &  0.896  &  1.031  &  2.335  &  0.635  &    --   &    --   &    --   &    --   \\
  4700  &  1.89  & &  0.952  &  1.074  &  2.443  &  0.667  &    --   &    --   &    --   &    --   \\
  4600  &  1.67  & &  1.014  &  1.123  &  2.561  &  0.701  &    --   &    --   &    --   &    --   \\
  4500  &  1.45  & &  1.084  &  1.177  &  2.691  &  0.737  &  1.091  &  1.147  &  2.674  &  0.741  \\
  4400  &  1.24  & &  1.164  &  1.241  &  2.832  &  0.775  &  1.170  &  1.207  &  2.810  &  0.778  \\
  4300  &  1.04  & &  1.248  &  1.310  &  2.987  &  0.817  &  1.252  &  1.273  &  2.957  &  0.819  \\
  4200  &  0.84  & &  1.342  &  1.390  &  3.159  &  0.860  &  1.338  &  1.343  &  3.117  &  0.863  \\
  4100  &  0.65  & &  1.440  &  1.479  &  3.351  &  0.910  &  1.425  &  1.418  &  3.289  &  0.911  \\
  4000  &  0.47  & &  1.543  &  1.579  &  3.557  &  0.963  &  1.514  &  1.501  &  3.475  &  0.964  \\
  3900  &  0.30  & &  1.656  &  1.695  &  3.789  &  1.018  &  1.606  &  1.592  &  3.680  &  1.022  \\
  3800  &  0.13  & &  1.774  &  1.824  &  4.043  &  1.078  &    --   &    --   &    --   &    --   \\
  3700  & --0.03 & &  1.896  &  1.968  &  4.316  &  1.139  &    --   &    --   &    --   &    --   \\
  3600  & --0.18 & &  2.017  &  2.150  &  4.631  &  1.196  &    --   &    --   &    --   &    --   \\
  3500  & --0.32 & &  1.998  &  2.243  &  4.941  &  1.319  &    --   &    --   &    --   &    --   \\

\noalign{\smallskip}

        &        & &                    \multicolumn{8}{c}{\FeoH=--2.0}                            \\

\noalign{\smallskip}

  4900  &  2.15  & &  0.820  &  1.002  &  2.240  &  0.604  &    --   &    --   &    --   &    --   \\
  4800  &  1.90  & &  0.878  &  1.046  &  2.341  &  0.632  &    --   &    --   &    --   &    --   \\
  4700  &  1.67  & &  0.944  &  1.094  &  2.453  &  0.662  &    --   &    --   &    --   &    --   \\
  4600  &  1.44  & &  1.018  &  1.148  &  2.574  &  0.692  &    --   &    --   &    --   &    --   \\
  4500  &  1.22  & &  1.107  &  1.214  &  2.711  &  0.725  &  1.094  &  1.183  &  2.706  &  0.738  \\
  4400  &  1.01  & &  1.200  &  1.285  &  2.858  &  0.759  &  1.186  &  1.253  &  2.849  &  0.773  \\
  4300  &  0.81  & &  1.304  &  1.370  &  3.029  &  0.798  &  1.282  &  1.327  &  3.006  &  0.813  \\
  4200  &  0.61  & &  1.417  &  1.465  &  3.218  &  0.842  &  1.382  &  1.406  &  3.175  &  0.857  \\
  4100  &  0.43  & &  1.534  &  1.571  &  3.426  &  0.889  &  1.484  &  1.490  &  3.358  &  0.906  \\
  4000  &  0.25  & &  1.660  &  1.696  &  3.664  &  0.939  &  1.588  &  1.580  &  3.555  &  0.961  \\
  3900  &  0.08  & &  1.787  &  1.828  &  3.918  &  0.994  &    --   &    --   &    --   &    --   \\
  3800  & --0.08 & &  1.917  &  1.981  &  4.212  &  1.059  &    --   &    --   &    --   &    --   \\
  3700  & --0.23 & &  2.050  &  2.150  &  4.524  &  1.121  &    --   &    --   &    --   &    --   \\
  3600  & --0.37 & &  2.178  &  2.328  &  4.843  &  1.181  &    --   &    --   &    --   &    --   \\
  3500  & --0.51 & &  2.207  &  2.568  &  5.140  &  1.175  &    --   &    --   &    --   &    --   \\

\noalign{\smallskip}
\hline

\end{tabular}
\end{table*}

It is worth remarking that the clusters in our sample are old;
thus the masses of their RGB stars should be low, typically
$\sim0.8-0.9$\,$M_{\odot}$ \citep{YY01}. Though the direct effect
of stellar mass on the broad-band photometric colors is small (see
Paper~I), $T_{\rm eff}$--$\log g$ relations will indeed be
different in case of younger stellar populations, because of the
higher masses of their RGB stars. For example, at $\MoH=-1.0$ and
$T_{\rm eff}=4400$\,K, the gravity on the 2\,Gyr isochrone will be
about 0.15 dex lower than $\log g$ corresponding to the same
effective temperature on the 15\,Gyr isochrone (this difference is
smaller at lower $T_{\rm eff}$ and $\MoH$, Yi et al. 2001).
Fortunately, the effect of this difference is small: a shift in
gravity by $\sim0.15$\,dex at $T_{\rm eff}=4400$\,K will translate
to $\Delta (B-V)\sim0.02$ for $\FeoH=-1.0$ and to $\sim0.03$ for
$\FeoH=-2.0$, with colors at lower $\log g$ becoming redder
(differences in other colors are smaller).

While the new $T_{\rm eff}$--$\log g$ relations provided in
Tables~\ref{tabGGCscales} and \ref{tabGGCscalesfits} cover the
effective temperature range of $T_{\rm eff}=3500-4900$\,K, it
should be taken into account that uncertainties in these relations
will likely be larger below $T_{\rm eff}\sim3900$\,K and above
$T_{\rm eff}\sim4600$\,K, where an extrapolation was used to
compensate for a lack of stars in certain cluster groups. It
should also be noted that the new $T_{\rm eff}$--$\log g$
relations provided in Tables~\ref{empGGCscales} and
\ref{tabGGCscales} are representative for RGB stars. Appropriate
care should thus be taken if these relations are used at higher
(or lower) temperatures, where RGB stars may be mixed up with
stars on the horizontal branch (or AGB stars).

\section{Synthetic photometric colors versus observations: results and discussion\label{teffscales}}

The new $T_{\rm eff}$--$\log g$ relations that were discussed in
the previous Section allow us to derive a set of new $T_{\rm
eff}$--$\log g$--color relations based on the synthetic
photometric colors, and to compare them with various $T_{\rm
eff}$--color and color--color relations available from the
literature. This comparison is done separately for $\MoH=-1.0$ and
$-2.0$. Below we focus on the details of these steps.

\subsection{New $T_{\rm eff}$--$\log g$--color scales}

The new $T_{\rm eff}$--$\log g$--color relations were constructed
using the $T_{\rm eff}$--$\log g$ relations derived in
Sect.~\ref{TGrelations} and broad-band photometric colors
calculated with the {\tt PHOENIX}, {\tt MARCS}, and {\tt ATLAS}
stellar model atmospheres (Sect.~\ref{synthcolors}). These
relations are provided in Table~\ref{TGPhoenixMarcs} (scales
employing {\tt PHOENIX} and {\tt MARCS} colors) and
Table~\ref{TGAtlas} (relation based on {\tt ATLAS} colors). They
are based on the new empirical $T_{\rm eff}$--$\log g$ scales
corresponding to the metallicity scale of KI03
(Sect.~\ref{TGrelations}), and are delivered at four
metallicities, $\MoH=-0.5$, $-1.0$, $-1.5$, and $-2.0$. Note that
the limiting gravities in {\tt MARCS} and {\tt ATLAS} grids of
synthetic colors are $\log g = 0.5$ and $\log g = 0.0$,
respectively; to extend the coverage in $\log g$, {\tt MARCS} and
{\tt ATLAS} colors were linearly extrapolated to $\sim0.2$\,dex
below these values.

Uncertainties in the new $T_{\rm eff}$--$\log g$--color relations
are governed by the uncertainties in the empirical $T_{\rm
eff}$--$\log g$ scales that were obtained as best fits to the
observed $T_{\rm eff}$--$\log g$ sequences of late-type giants in
Galactic globular clusters (Sect.~\ref{TGrelations}). The typical
RMS residual of the fitting procedure is $\simeq0.15$\,dex in
$\log g$, or $\simeq100$\,K in $T_{\rm eff}$. At $T_{\rm
eff}=4400$\,K, $\log g=1.5$, and $\MoH=-1.0$, the uncertainty
$\Delta T_{\rm eff}=100$\,K will be equivalent to changes in
photometric colors $\Delta(B-V)\simeq\Delta(V-I)\simeq0.05$,
$\Delta(V-K)\simeq0.13$, and $\Delta(J-K)\simeq0.04$
(correspondingly, to 0.07, 0.05, 0.14, and 0.04\,mag, at $\log
g=1.0$ and $\MoH=-2.0$). The effect on photometric colors will
increase slightly with decreasing gravity. Note, however, that
photometric colors are less sensitive to uncertainties in $\log
g$: at $T_{\rm eff}=4400$\,K, $\log g=1.5$, and $\MoH=-1.0$,
$\Delta \log g =0.15$ will correspond to $\Delta(B-V)\simeq0.02$,
with differences in other colors at the level of 0.01\,mag or
lower. While we will further quote $\pm100$\,K as a representative
uncertainty of the new $T_{\rm eff}$--$\log g$--color relations,
it is rather obvious that this may only represent a lower limit on
the true uncertainties, which include various systematical effects
inherent in the spectroscopic derivations of $T_{\rm eff}$, $\log
g$, and $\FeoH$, limitations of the current stellar atmosphere
models, and so forth.

\begin{table}[t]
\caption[]{$T_{\rm eff}$--$\log g$--color relations for the
late-type giants based on the synthetic colors calculated with
{\tt ATLAS} model atmospheres. Photometric colors are given in the
Johnson-Cousins-Glass system (see Sect.~\ref{synthcolors} for
details). \label{TGAtlas}}

\centering \scriptsize

\begin{tabular}{ccccccccccc}
\hline \noalign{\smallskip}
        &       & &          \multicolumn{4}{c}{ATLAS}    \\
\cline{4-7} \noalign{\smallskip}
 $T_{\rm eff}$  & $\log g$ &  & $B\!-\!V$ & $V\!-\!I$ & $V\!-\!K$ & $J\!-\!K$ \\
\noalign{\smallskip} \hline \noalign{\smallskip}

              \multicolumn{7}{c}{\FeoH=--0.5}             \\

\noalign{\smallskip}

  4750  &  2.50  & &  1.010  &  1.033  &  2.383  &  0.660  \\
  4500  &  1.96  & &  1.133  &  1.141  &  2.671  &  0.751  \\
  4250  &  1.47  & &  1.273  &  1.275  &  3.013  &  0.857  \\
  4000  &  1.00  & &  1.453  &  1.465  &  3.438  &  0.974  \\
  3750  &  0.56  & &  1.657  &  1.739  &  3.975  &  1.098  \\
  3500  &  0.17  & &  1.811  &  2.176  &  4.695  &  1.196  \\

\noalign{\smallskip}

              \multicolumn{7}{c}{\FeoH=--1.0}             \\

\noalign{\smallskip}

  4750  &  2.24  & &  0.959  &  1.030  &  2.386  &  0.666  \\
  4500  &  1.68  & &  1.098  &  1.142  &  2.673  &  0.755  \\
  4250  &  1.18  & &  1.259  &  1.283  &  3.019  &  0.859  \\
  4000  &  0.72  & &  1.465  &  1.478  &  3.453  &  0.978  \\
  3750  &  0.31  & &  1.708  &  1.747  &  3.989  &  1.107  \\
  3500  & --0.07 & &  1.946  &  2.109  &  4.606  &  1.206  \\

\noalign{\smallskip}

              \multicolumn{7}{c}{\FeoH=--1.5}             \\

\noalign{\smallskip}

  4750  &  2.01  & &  0.926  &  1.037  &  2.395  &  0.668  \\
  4500  &  1.45  & &  1.089  &  1.161  &  2.688  &  0.753  \\
  4250  &  0.94  & &  1.285  &  1.322  &  3.051  &  0.853  \\
  4000  &  0.47  & &  1.532  &  1.542  &  3.512  &  0.971  \\
  3750  &  0.05  & &  1.813  &  1.839  &  4.090  &  1.104  \\
  3500  & --0.32 & &    --   &    --   &    --   &    --   \\

\noalign{\smallskip}

              \multicolumn{7}{c}{\FeoH=--2.0}             \\

\noalign{\smallskip}

  4750  &  1.79  & &  0.900  &  1.052  &  2.408  &  0.667  \\
  4500  &  1.22  & &  1.099  &  1.198  &  2.715  &  0.745  \\
  4250  &  0.71  & &  1.344  &  1.397  &  3.111  &  0.838  \\
  4000  &  0.25  & &  1.636  &  1.666  &  3.625  &  0.952  \\
  3750  & --0.16 & &    --   &    --   &    --   &    --   \\
  3500  & --0.51 & &    --   &    --   &    --   &    --   \\

\noalign{\smallskip} \hline

\end{tabular}
\end{table}

\subsection{Comparison of $T_{\rm eff}$--color and color--color relations\label{compar}}

In principle, the new $T_{\rm eff}$--$\log g$--color relations can
be readily used to compare synthetic photometric colors with
observed effective temperatures and colors of late-type giants in
the $T_{\rm eff}$--color planes. Such an approach was taken in
Paper~I, where for this purpose we used a sample of late-type
giants with effective temperatures derived from interferometric
measurements of stellar radii. Unfortunately, as was already
mentioned in Sect.~\ref{TGsample}, effective temperatures of
late-type giants obtained from interferometry or lunar
occultations are very scarce at sub-Solar metallicities. The
sample of late-type giants in Galactic globular clusters which was
employed in the previous Section to obtain the new $T_{\rm
eff}$--$\log g$ relations can not be used for this purpose either,
since the number of stars in each metallicity group is too small
for a reliable comparison of observed and synthetic photometric
colors in different $T_{\rm eff}$--color planes at different
metallicities.

Instead of using effective temperatures and photometric colors of
individual stars we thus will make a comparison of synthetic
photometric colors with the $T_{\rm eff}$--color and color--color
relations available from the literature. For this purpose we use
relations based both on the observed and theoretical colors of
late-type giants, at $\MoH=-1.0$ and $-2.0$.

The baseline set of $T_{\rm eff}$--color and color--color
relations used in our comparisons is that of \citet[][ A99]{A99}.
These relations are built on the observed properties of a
homogenous sample of 250 late-type giants in Galactic globular
clusters, with precise optical and near-infrared photometry of
individual stars and $T_{\rm eff}$ from IRFM. The sample covers
the metallicity range $\FeoH=0\dots-3.0$, with $\FeoH$ estimates
obtained either from spectroscopy or Str\"{o}mgren photometry
(with typical accuracies of $\pm0.15$\,dex and $0.2$ to
$0.3$\,dex, respectively). The IRFM temperatures of individual
stars are typically in good agreement with those obtained by
direct methods within a large range of effective temperatures
($T_{\rm eff}\sim3700\dots5200$\,K). The mean difference between
the interferometric and IRFM temperatures is $T_{\rm
direct}-T_{\rm IRFM}=3\pm51$\,K, based on 20 stars \citep{AAM99}.
The internal accuracy of the individual $T_{\rm eff}$--color
relations varies between $40$ and $125$\,K, which compares well
with the accuracies of the $T_{\rm eff}$--color scales derived by
us in Paper~I employing a sample of stars with interferometric
temperatures (which are of the order of $\pm150$\,K). Before
making the comparisons, the $T_{\rm eff}$--color relations of A99
(Table~6 of A99, Johnson system) were transformed to the standard
Johnson-Cousins-Glass system (Sect.~\ref{synthcolors}), using
transformation equations from \citet{F83} for $(V-I)$, and from
\citet{BB88} for $(V-K)$ and $(J-K)$.

Recently, the $T_{\rm eff}$--color and color--color relations of
A99 were updated by \citet[][]{RM05a, RM05b}. We include these new
relations into our analysis, too. The transformation of \citet[][
RM05]{RM05b} colors involving the 2MASS bandpasses ($V-K_2$,
$J_2-K_2$; subscript 2 denotes the 2MASS system) to the standard
Johnson-Cousins-Glass system was made using transformation
equations given by \citet{C01}.

Several additional widely used $T_{\rm eff}$--color and
color--color relations that were employed in our study are:

\begin{itemize}

\item {\em \citet[][ BG89]{BG89}:} $T_{\rm eff}$--color and
color--color relations based on theoretical photometric colors.
The BG89 scales employed in this study were constructed using
$B-V$ colors taken from \citet{BG78}, $V-I$, $V-K$, and $J-K$
colors from \citet{BG89}. According to \citet[][]{VC03}, the scale
of BG89 reproduces well the observed CMDs of Galactic open and
globular clusters at different metallicities. In the effective
temperature range of interest for this study, the agreement seems
to be very good, both at $\MoH=-1.0$ and $-2.0$ \citep[][]{VC03};

\item {\em BaSeL 2.2 \citep[][ BaSeL 2.2]{LCB98}:} a semiempirical
library of photometric colors based on the theoretical spectra.
While photometric colors of BaSeL 2.2 are calibrated to match
empirical $T_{\rm eff}$--color relations at $\FeoH=0.0$, with
presumably poorer consistency at lower metallicities, we employ
this scale in the present study too, partially to compare it with
the BaSeL 3.1 colors (see below). BaSeL 2.2 colors used in this
work were calculated using the interactive web-based BaSeL
server\footnote{http://tangerine.astro.mat.uc.pt/BaSeL/};

\item {\em BaSeL 3.1 \citep[][ BaSeL 3.1]{WLB02}:} extension of the
BaSeL 2.2 library to lower metallicities, calibrated using the
photometric data of Galactic globular clusters. This scale is
designed to reproduce the $T_{\rm eff}$--color and color-color
relations at metallicities down to $\MoH\sim-2.0$. BaSeL 3.1
colors used in this study were calculated using the web-based
BaSeL server;

\item {\em \citet[][ H00]{H00}:} synthetic colors based on theoretical
spectra calculated with the MARCS and SSG codes, with TiO opacities
adjusted to reproduce the observed spectra of M giants from
\citet{F94}. The H00 scale should be used with care below $T_{\rm
eff}\sim3800$\,K since no ${\rm H_2O}$ opacities were included in
the calculations (see Paper~I for more details on the influence of
different molecular opacities on the photometric colors);

\item {\em \citet[][ SF00]{SF00}:} an empirical $T_{\rm eff}$--$(B-V)$
scale based on the observed colors and effective temperatures of
537 {\it Infrared Space Observatory} (ISO) standard stars from
\citet{DB98}. Effective temperatures of individual stars were
derived from the $T_{\rm eff}$--$(V-K)$ relation, calibrated on a
sample of nearby stars with angular diameters available from
interferometry;

\item {\em \citet[][ VC03]{VC03}:} empirical scales based on
synthetic \emph{BVRI} colors of \citet{BG78,BG89}, adjusted to
satisfy observational constrains from the observed CMDs of
Galactic globular and open clusters, field stars in the Solar
neighborhood, empirical $T_{\rm eff}$--color relations and
color--color relations for field giants.

\end{itemize}

Note that in all cases photometric colors were selected according
to the new $T_{\rm eff}$--$\log g$ relations derived in
Sect.~\ref{TGrelations}.

Extensive comparisons of these relations with various other
$T_{\rm eff}$--color scales have been published in numerous
studies \citep[e.g.,][]{A99, H00, SF00, WLB02, VC03, RM05b}. Some
of these relations (BaSeL 2.2, A99, H00, SF00, and VC03) were also
employed in our comparison of the observed and theoretical colors
of late-type giants at Solar metallicity (Paper~I). Altogether,
this provides a further reference list for a comparison of the new
$T_{\rm eff}$--color and color--color scales with similar
relations available in the literature.

Comparisons of the new $T_{\rm eff}$--$\log g$--color relations
based on {\tt PHOENIX}, {\tt MARCS} and {\tt ATLAS} colors
(Tables~\ref{TGPhoenixMarcs} and \ref{TGAtlas}) with the published
$T_{\rm eff}$--color relations are given in
Fig.~\ref{figTCplanes-1.0} ($T_{\rm eff}$--color planes) and
Fig.~\ref{figCCplanes-1.0} (color--color planes) at $\MoH=-1.0$,
and, correspondingly, in Figs.~\ref{figTCplanes-2.0} and
\ref{figCCplanes-2.0} at $\MoH=-2.0$ (all comparisons are made
with respect to the scale of A99). We discuss the trends at these
two metalicities in the sections below.

\subsubsection{$\MoH=-1.0$}

\begin{figure*}[p]
\centering
\includegraphics[width=16.6cm] {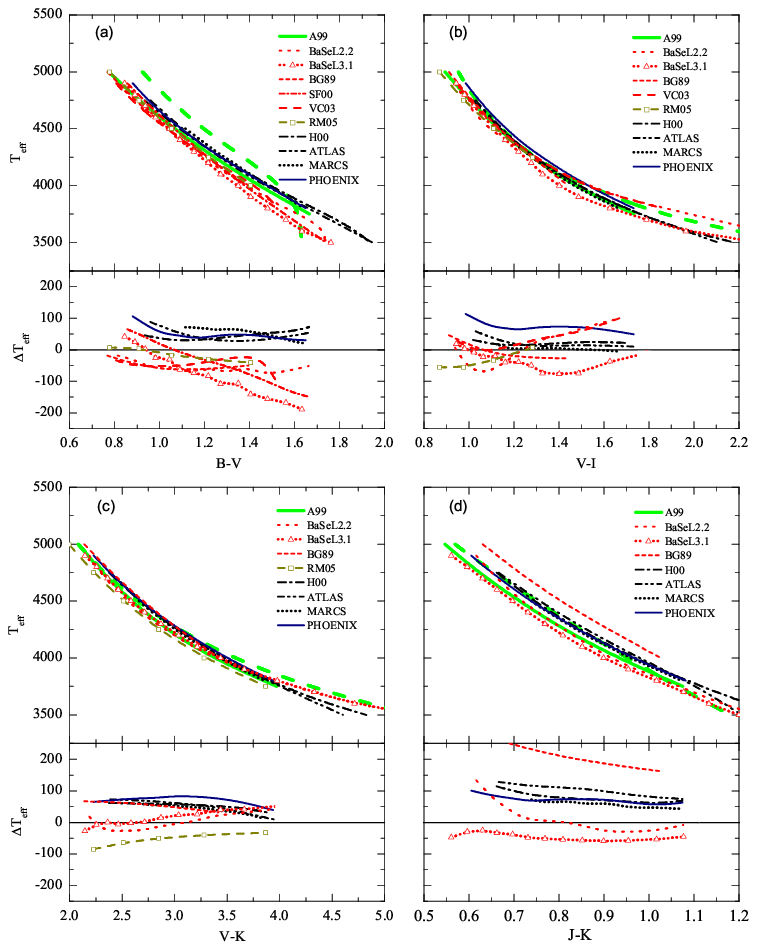}
\caption[]{Empirical and theoretical $T_{\rm eff}$--color
relations for late-type giants at $\MoH=-1.0$, in different
$T_{\rm eff}$--color planes (a-d, top panels). The thick solid
line shows $T_{\rm eff}$--color relation of A99, the thick dashed
line is the empirical $T_{\rm eff}$--color relation at Solar
metallicity (Paper~I). Several existing $T_{\rm eff}$--color
relations are shown as well, together with the scales constructed
using synthetic colors of {\tt PHOENIX}, {\tt MARCS} and {\tt
ATLAS} (Tables~\ref{TGPhoenixMarcs} and \ref{TGAtlas}). The bottom
panels in each figure show the difference between various $T_{\rm
eff}$--color relations and the A99 scale in a given $T_{\rm
eff}$--color plane ($\Delta T_{\rm eff}=T_{\rm eff}^{other}-T_{\rm
eff}^{A99}$.)\label{figTCplanes-1.0}}
\end{figure*}

\begin{figure}[t]
\centering
\includegraphics[width=8.5cm] {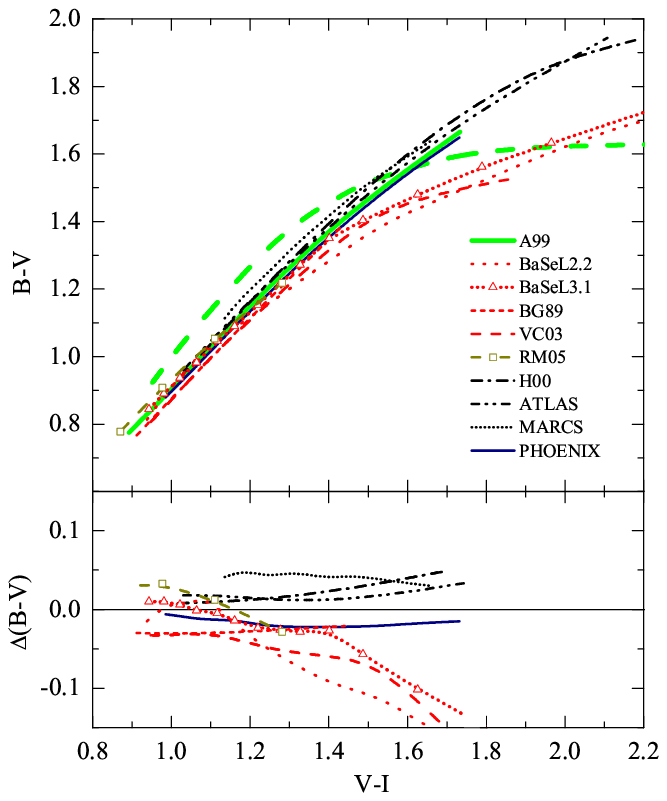}
\caption[]{(a) {\bf Top}: Empirical and theoretical color--color
relations at $\MoH=-1.0$ in the $(B-V)-(V-I)$ plane. The thick
solid line is the $(B-V)-(V-I)$ relation of A99, the thick dashed
line is the empirical $T_{\rm eff}$--color relation at Solar
metallicity (Paper~I). Several existing $T_{\rm eff}$--color
relations are shown, together with the scales constructed using
synthetic colors of {\tt PHOENIX}, {\tt MARCS} and {\tt ATLAS}.
{\bf Bottom:} the difference between various $(B-V)-(V-I)$
relations and the A99 scale, $\Delta {\rm CI}={\rm
CI}^{other}-{\rm CI}^{A99}$.\label{figCCplanes-1.0}}
\end{figure}

\begin{figure}[tb]
\centering \addtocounter{figure}{-1}
\includegraphics[width=8.5cm] {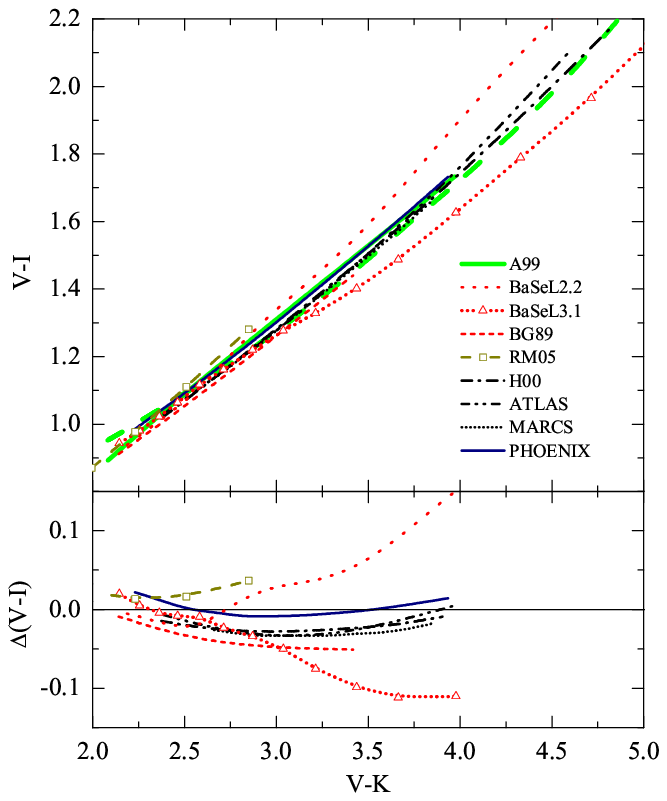}
\caption[]{(b) Same as in Fig.~\ref{figCCplanes-1.0}a but in the
$(V-I)-(V-K)$ plane.}
\end{figure}

\begin{figure}[tb]
\centering \addtocounter{figure}{-1}
\includegraphics[width=8.5cm] {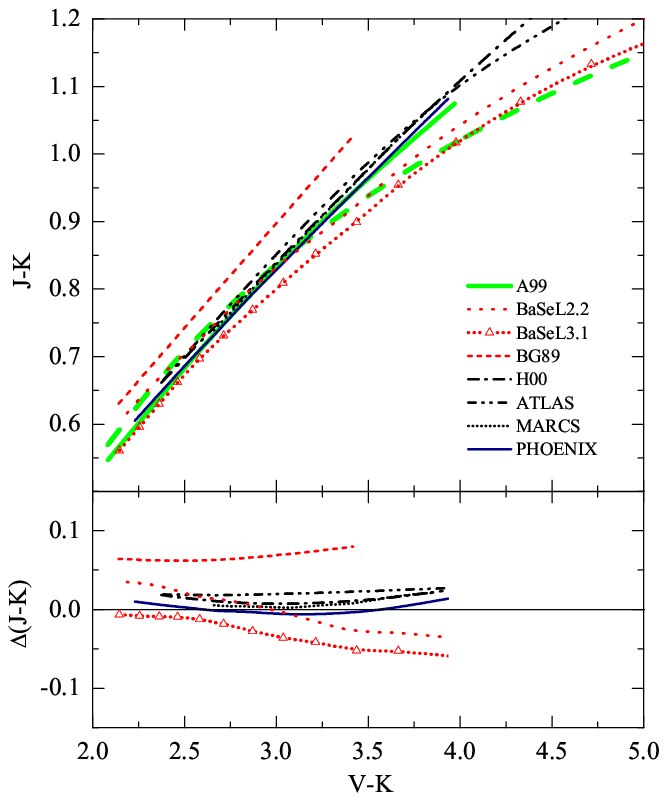}
\caption[]{(c) Same as in  Fig.~\ref{figCCplanes-1.0}a but in the
$(J-K)-(V-K)$ plane.}
\end{figure}

On the whole, the agreement between different $T_{\rm eff}$--color
relations is good in all $T_{\rm eff}$--color planes
(Fig.~\ref{figTCplanes-1.0}). Typical differences are well within
$\Delta T_{\rm eff} \sim80$\,K, with somewhat larger deviations in
the $T_{\rm eff}$--$(J-K)$ plane. Except for the scales of SF00
and BaSeL 3.1 (which start to deviate below $T_{\rm
eff}\sim4100$\,K and $\sim4300$\,K, correspondingly), reasonably
good agreement is also seen in the $T_{\rm eff}$--$(B-V)$ plane
(note that effective temperature--color relations differ
considerably more in this plane at Solar metallicity; see
Paper~I). It is worthwhile noting that the slopes of SF00 and
BaSeL 3.1 scales are clearly different from those of other
relations in this $T_{\rm eff}$--color plane. The agreement
between different relations is very good in the $T_{\rm
eff}$--$(V-I)$ plane, typically to $\pm50$\,K, with somewhat
larger deviations for the scales based on BaSeL and {\tt PHOENIX}
colors. There is also a good consistency between different $T_{\rm
eff}$--color scales in the $T_{\rm eff}$--$(V-K)$ plane (to
$\pm80$\,K), though effective temperatures predicted by the
$T_{\rm eff}$--color relation of A99 are slightly lower than those
resulting from other relations.

In spite of the reasonably good agreement between the different
$T_{\rm eff}$--color scales in general, there is a clear
indication that effective temperatures predicted by the $T_{\rm
eff}$--color relations based on synthetic colors (H00, {\tt
ATLAS}, {\tt MARCS}, {\tt PHOENIX}) are typically slightly higher
than those inferred from the empirical relations (i.e., those of
A99, BaSeL 2.2 and 3.1, SF00, VC03, RM05). This offset is seen in
all $T_{\rm eff}$--color planes and is largest in the $T_{\rm
eff}$--$(B-V)$ plane, where the average difference between the
predictions of theoretical and empirical relations amounts to
$\sim100$\,K (similar offset is seen in the $T_{\rm eff}$--$(J-K)$
plane too, though in this case the comparison can only be made
with the semi-empirical BaSel 2.2 and 3.1 scales). The only
exception in this sense is the scale of BG89: theoretical colors
of BG89 are very similar to those predicted by empirical relations
in the $T_{\rm eff}$--$(B-V)$ and $T_{\rm eff}$--$(V-I)$ planes.
In the $T_{\rm eff}$--$(V-K)$ plane, the BG89 relation is similar
to the other scales based on synthetic colors, while in the
$T_{\rm eff}$--$(J-K)$ plane it predicts effective temperatures
that are considerably higher than those resulting from other
$T_{\rm eff}$--color relations.

It should be noted that the A99 scale occupies an intermediate
position in this sense in all $T_{\rm eff}$--color planes,
providing a compromise between the predictions of theoretical and
empirical $T_{\rm eff}$--color relations. That is, effective
temperatures given by the A99 scale generally tend to be lower
than those given by $T_{\rm eff}$--color relations based on
synthetic colors, with an average offset of about $\sim60$\,K. On
the other hand, they are higher by up to $\sim50$\,K ($T_{\rm
eff}$--$(B-V)$ plane) than effective temperatures predicted by the
empirical scales. \citet{I01} have reached similar conclusions in
their comparison of the effective temperatures of RGB stars in the
globular cluster M5 derived using the $T_{\rm eff}$--$(B-V)$
scales of A99, SF00 and H00. While there are some hints that the
A99 scale tends to predict slightly lower $T_{\rm eff}$ than other
$T_{\rm eff}$--color relations at Solar metallicity (Paper~I), at
$\MoH=-1.0$ a similar trend is seen only in the $T_{\rm
eff}$--$(V-K)$ plane (it is interesting to note in this respect
that the $T_{\rm eff}$'s predicted by A99 scales at $\MoH=-2.0$
are indeed lower than those resulting from other $T_{\rm
eff}$--color relations; see Sect.~\ref{MoH=-2.0}).

While the RM05 scale (which is an extension and update of the A99
relations) predicts slightly different effective temperatures than
do the A99 relations, differences between the two scales are
always within $\pm50$\,K. A slightly larger discrepancy is seen in
the $T_{\rm eff}$--$(V-K)$ plane, especially at $T_{\rm
eff}\ga4300$\,K. Note, however, that the transformation equations
of \citet{C01} used by us to convert $(V-K)_{\rm 2MASS}$ colors of
RM05 to the Johnson-Cousins-Glass system (see
Sect.~\ref{synthcolors} for details) were derived utilizing only a
part of the 2MASS survey data available at that time.
Nevertheless, a shift of $\Delta (V-K)=0.1$ is needed to
compensate for a difference of $50$\,K between the A99 and RM05
relations (Fig.~\ref{figTCplanes-1.0}c), which is quite large to
be explained by uncertainties in the transformation equations.

We find no clear evidence that the $T_{\rm eff}$--color relations
based on the BaSeL 3.1 colors would indeed represent an
improvement over the BaSeL 2.2 scales at $\MoH=-1.0$. In fact, the
scales employing BaSeL 2.2 colors seem to be in better agreement
with the general trends seen in the $T_{\rm eff}$--$(B-V)$ and
$T_{\rm eff}$--$(V-I)$ planes at $T_{\rm eff}\la4500$\,K.
Otherwise, the two scales behave very similarly, especially in the
$T_{\rm eff}$--$(V-K)$ and $T_{\rm eff}$--$(J-K)$ planes (though
differences gradually start to build up in the latter case at
$T_{\rm eff}>4500$\,K).

The agreement between synthetic colors calculated using different
stellar atmosphere models is very good (including the scale based
on H00 colors), with differences typically well within $\Delta
T_{\rm eff}\sim70$\,K. A somewhat larger discrepancies are seen in
the $T_{\rm eff}$--$(V-I)$ plane, where the $T_{\rm eff}$--color
scale based on the {\tt PHOENIX} colors tends to predict somewhat
higher effective temperatures than those resulting from other
$T_{\rm eff}$--color relations (a similar trend is seen at Solar
metallicity, see Paper~I).

The $T_{\rm eff}$--color relations of BG89 are based on
theoretical colors that were calculated nearly 20 years ago
\citep{BG78, BG89}. This offers an intriguing possibility for
examining the changes/improvements made in the theoretical
modeling of stellar spectra and photometric colors over the last
two decades, by comparing the predictions of BG89 with the new
$T_{\rm eff}$--color relations based on colors calculated with the
current state-of-the-art stellar model atmospheres (i.e., {\tt
PHOENIX}, {\tt MARCS}, and {\tt ATLAS}). Interestingly, very
little difference is seen between the two sets of theoretical
$T_{\rm eff}$--color relations in the $T_{\rm eff}$--$(V-I)$ and
$T_{\rm eff}$--$(V-K)$ planes. This is remarkable, especially
since the $V$, $I$, and $K$ passbands are strongly influenced by
various molecular bands (TiO, VO, ${\rm H}_{2}{\rm O}$, etc.). The
differences are larger in the $T_{\rm eff}$--$(B-V)$ and $T_{\rm
eff}$--$(J-K)$ planes though. Compared with the effective
temperatures obtained using $T_{\rm eff}$--color relations based
on the synthetic colors of {\tt PHOENIX}, {\tt MARCS}, and {\tt
ATLAS}, the $T_{\rm eff}$ predicted by the BG89 scale are
approximately $\sim100$\,K lower in the former case and by a
similar amount higher in the latter. On the whole, the differences
in photometric colors provided by BG89 and those calculated with
the current stellar model atmospheres are not large, typically
$\la100$\,K.

The agreement between different color--color relations is good in
all color--color planes (Fig.~\ref{figCCplanes-1.0}). The
differences are well within $\sim\pm0.05$ mag in the
$(B-V)$--$(V-I)$ plane, while the agreement is even better in the
$(V-I)$--$(V-K)$ and $(J-K)$--$(V-K)$ planes (to $\sim\pm0.03$
mag). Note, however, that the scales based on BaSeL 2.2. and 3.1
colors are rather deviant in all color--color planes at the redder
colors (below $T_{\rm eff}\sim4000-4300$\,K). Similarly, the VC03
relation starts to deviate beyond $V-I\simeq1.5$ in the
$(B-V)$--$(V-I)$ plane. BG89 colors are discrepant in the
$(J-K)$--$(V-K)$ plane (due to deviations in the $T_{\rm
eff}$--$(J-K)$ plane), within the entire range of photometric
colors (or effective temperatures) typical for late-type giants.

\subsubsection{$\MoH=-2.0$\label{MoH=-2.0}}

\begin{figure*}[p]
\centering
\includegraphics[width=16.6cm] {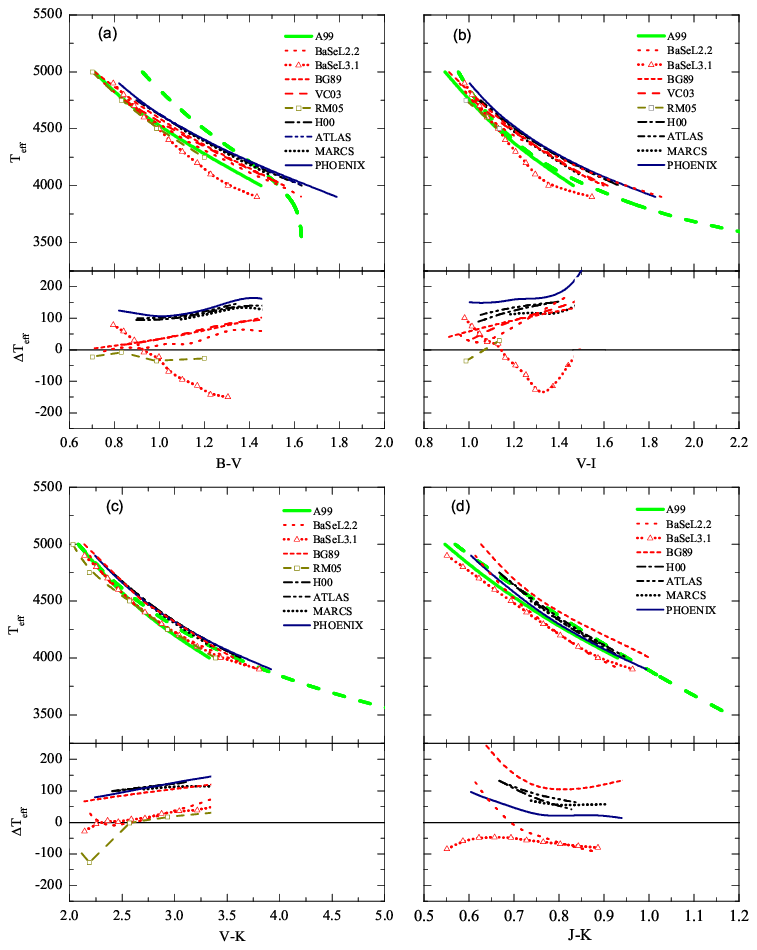}
\caption[]{Empirical and theoretical $T_{\rm eff}$--color
relations at $\FeoH=-2.0$, in different $T_{\rm eff}$--color
planes (a-d, top panels). The thick solid line is the $T_{\rm
eff}$--color relation of A99, the thick dashed line is empirical
$T_{\rm eff}$--color relation at Solar metallicity (Paper~I).
Several existing $T_{\rm eff}$--color relations are shown as well,
together with the scales constructed using synthetic colors of
{\tt PHOENIX}, {\tt MARCS} and {\tt ATLAS}
(Tables~\ref{TGPhoenixMarcs} and \ref{TGAtlas}). The bottom panels
in each figure show the difference between various $T_{\rm
eff}$--color relations and the A99 scale in a given $T_{\rm
eff}$--color plane ($\Delta T_{\rm eff}=T_{\rm eff}^{other}-T_{\rm
eff}^{A99}$.)\label{figTCplanes-2.0}}
\end{figure*}

\begin{figure}[t]
\centering
\includegraphics[width=8.5cm] {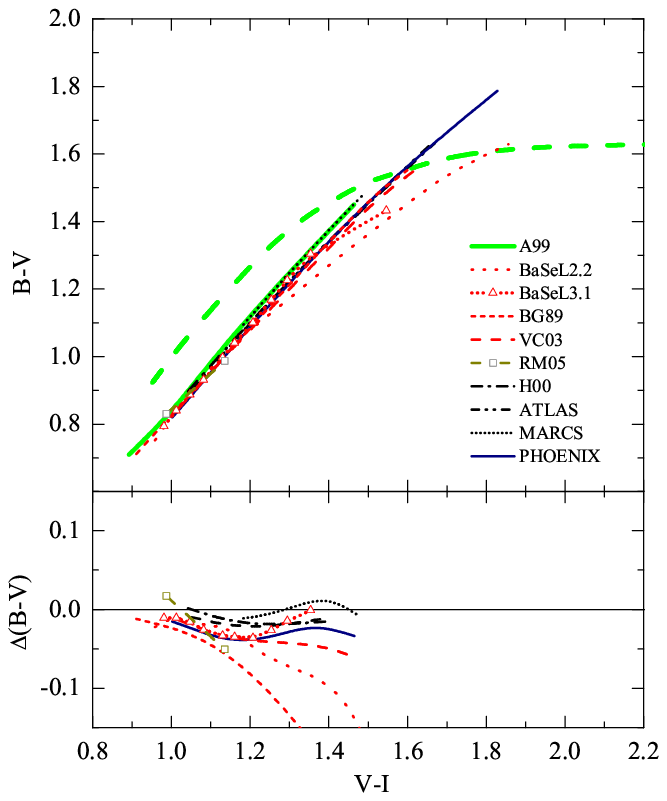}
\caption[]{(a) {\bf Top}: Empirical and theoretical color--color
relations at $\FeoH=-2.0$, in the $(B-V)-(V-I)$ plane. The thick
solid line is the $(B-V)-(V-I)$ relation of A99, the thick dashed
line is empirical $T_{\rm eff}$--color relation at Solar
metallicity (Paper~I). Several existing $T_{\rm eff}$--color
relations are shown, together with the scales constructed using
synthetic colors of {\tt PHOENIX}, {\tt MARCS} and {\tt ATLAS}.
{\bf Bottom:} the difference between various $(B-V)-(V-I)$
relations and the A99 scale, $\Delta {\rm CI}={\rm
CI}^{other}-{\rm CI}^{A99}$.\label{figCCplanes-2.0}}
\end{figure}

\begin{figure}[t]
\centering \addtocounter{figure}{-1}
\includegraphics[width=8.5cm] {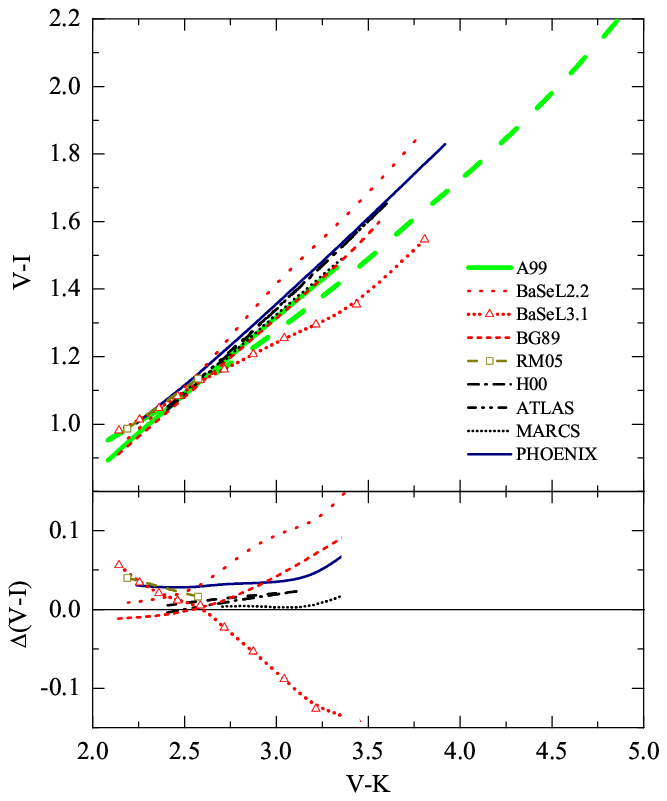}
\caption[]{(b) Same as in Fig.~\ref{figCCplanes-2.0}a but in the
$(V-I)-(V-K)$ plane.}
\end{figure}

\begin{figure}[tb]
\centering \addtocounter{figure}{-1}
\includegraphics[width=8.5cm] {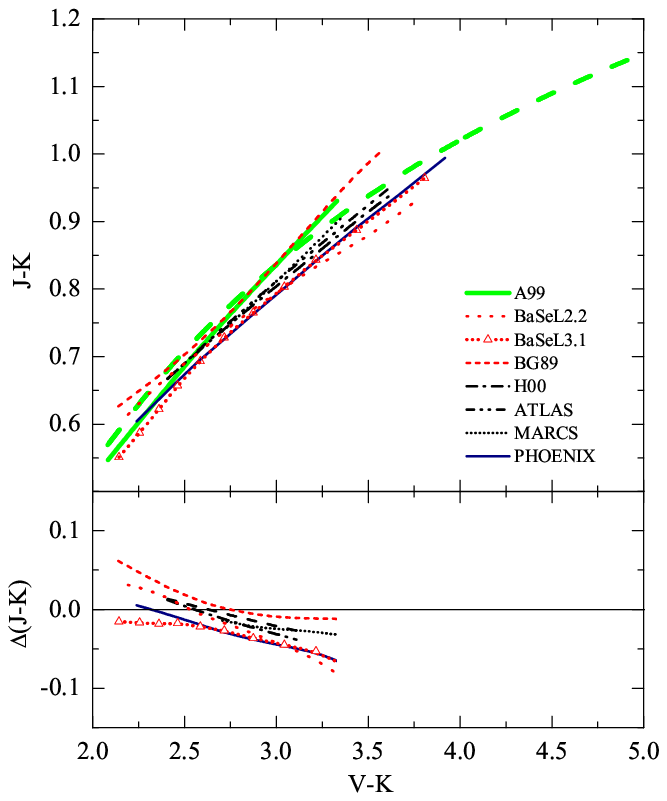}
\caption[]{(c) Same as in Fig.~\ref{figCCplanes-2.0}a but in the
$(J-K)-(V-K)$ plane.}
\end{figure}

Obviously, the agreement between various $T_{\rm eff}$--color
relations and the scale of A99 is poorer at $\FeoH=-2.0$
(Fig.~\ref{figTCplanes-2.0}). Effective temperatures predicted by
the A99 scales are systematically lower than those resulting from
the other $T_{\rm eff}$--color relations, by up to $\sim150$\,K.
This tendency is clearly seen in all $T_{\rm eff}$--color planes,
and applies to scales based both on observed and theoretical
colors (though the discrepancies are typically larger in the
latter case, to $\sim 130$\,K on average). It should be noted,
however, that A99 relations are in reasonable agreement with other
empirical relations in the $T_{\rm eff}$--$(B-V)$ and $T_{\rm
eff}$--$(V-K)$ planes (to $\sim90$\,K and $\sim70K$,
respectively); larger deviations occur only when A99 relations are
compared with those employing theoretical colors. Whether these
inadequacies point out to deficiencies in the A99 calibrations, or
problems with the predictions of theoretical models (or both) -
this has to be clarified in the future studies.

The agreement between $T_{\rm eff}$--color scales based on
different sets of synthetic colors is very good, with deviations
typically within $\Delta T_{\rm eff}\sim40$\,K. This is not an
unexpected result, since differences in the setup of stellar
atmosphere models (opacities, equation of state, etc.) generally
become less important at lower metallicities as their effect on
the resulting model structures becomes smaller too. Scales based
on {\tt PHOENIX} colors are slightly more discrepant in the
$T_{\rm eff}$--$(V-I)$ and $T_{\rm eff}$--$(J-K)$ planes, though
in the latter case this results in a better agreement with the A99
scale. The $T_{\rm eff}$--color relations based on BG89 colors are
very similar to those employing theoretical colors calculated with
current {\tt PHOENIX}, {\tt MARCS}, and {\tt ATLAS} stellar model
atmospheres. Significant differences are seen only in the $T_{\rm
eff}$--$(B-V)$ plane, where effective temperatures predicted by
the BG89 relations are cooler by about $\sim100$\,K, and in the
$T_{\rm eff}$--$(J-K)$ plane, where BG89 scales predict effective
temperatures that are systematically higher than those resulting
from the other scales based on synthetic colors.

Similarly to what was seen at $\MoH=-1.0$, we find no clear
evidence that scales employing BaSeL 3.1 colors would indeed
perform better than those based on BaSeL 2.2. In fact, BaSeL 2.2
colors are in much better agreement with the general trends seen
in the $T_{\rm eff}$--$(B-V)$ and $T_{\rm eff}$--$(V-I)$ planes,
both with respect to the scale of A99 and other $T_{\rm
eff}$--color relations. On the contrary, BaSeL 3.1 scales deviate
rapidly below $T_{\rm eff}\sim4600$\,K in the $T_{\rm
eff}$--$(B-V)$ and $T_{\rm eff}$--$(V-I)$ planes, predicting
effective temperatures that are too low. The two sets of relations
are nearly identical in the $T_{\rm eff}$--$(V-K)$ plane, with
slightly larger differences seen in the $T_{\rm eff}$--$(J-K)$
plane at $T_{\rm eff} \ga 4500$\,K.

The new scales of RM05 behave very similarly to those of A99. The
differences are always within $\sim \pm40$\,K, except that RM05
temperatures in the $T_{\rm eff}$--$(V-K)$ plane are up to
$\sim120$\,K lower than those predicted by empirical scales at
$\ga4500$\,K, and to $\sim200$\,K lower than those resulting from
the relations based on theoretical colors.

The agreement between different scales in the color--color planes
is generally very good. The differences are typically well within
$\sim\pm0.03$\,mag (except for scales based on BaSeL colors), with
a marginally larger scatter seen in the $(J-K)$--$(V-K)$ plane,
$\sim\pm0.05$\,mag. The only outliers are scales based on BaSeL
2.2 and 3.1 colors, as they deviate rapidly in the
$(B-V)$--$(V-I)$ and $(V-I)$--$(V-K)$ planes beyond
$(V-I)\simeq1.2$ and $(V-K)\simeq2.6$.

Indeed, most of the $T_{\rm eff}$--color scales employed in this
comparison are sensitive to the $T_{\rm eff}$--$\log g$ relations
used in their derivation. However, uncertainties in the $T_{\rm
eff}$--$\log g$ relations cannot fully account for the size of
systematic offsets seen above, e.g., those between the A99
relations and other $T_{\rm eff}$--color scales, as these
differences are simply too large. Theoretical models predict that
at a given photometric color higher effective temperatures will
correspond to the models with lower gravities. Our comparison
shows that effective temperatures inferred from the different
$T_{\rm eff}$--color relations are indeed always higher than those
resulting from the A99 scales. Obviously, such differences may
result if gravities predicted by the new $T_{\rm eff}$--$\log g$
relations are systematically too low. However, in order to account
for the average offset of $\sim100$\,K seen in the $T_{\rm
eff}$--$(V-I)$ and $T_{\rm eff}$--$(V-K)$ planes, gravities in the
$T_{\rm eff}$--$\log g$ relations should be increased
correspondingly by about 0.5 and 1\,dex at $\MoH=-2.0$ (generally,
larger changes will be needed at higher metallicities). A shift of
$\sim0.4$\,dex in gravity will be required to remedy the situation
in the $T_{\rm eff}$--$(B-V)$ plane. Clearly, the required shifts
in gravity are too large to offer a plausible solution in this
situation, especially given the fact that the gravity difference
between the empirical $T_{\rm eff}$--$\log g$ relations
(Sect.~\ref{TGrelations}) at $\MoH=-1.0$ and $-2.0$ is only about
$\sim0.5$\,dex.

\section{Conclusions}

A detailed investigation of the metallicity effects on the
broad-band photometric colors of late-type giants shows that their
photometric properties are generally little affected by the
variations in $\MoH$ at effective temperatures higher than $T_{\rm
eff}\sim3800$\,K. This picture gradually changes at lower $T_{\rm
eff}$, as the influence of metallicity becomes more strongly
pronounced below $T_{\rm eff}<3500$\,K. To a large extent this is
due to changing efficiency of molecule formation with
decreasing $\MoH$: since spectra of the late-type giants are
heavily blanketed by molecular lines (especially at low $T_{\rm
eff}$), photometric colors are inevitably affected when molecular
lines become weaker at lower $\MoH$.

In order to compare synthetic colors with observations of
late-type giants, we derive a set of new $T_{\rm eff}$--$\log
g$--color scales based on spectroscopic and photometric effective
temperatures and gravities of 152 late-type giants in 10 Galactic
globular clusters, backed up with synthetic colors produced with
{\tt PHOENIX}, {\tt MARCS} and {\tt ATLAS} stellar model
atmospheres. We find that the new scales employing synthetic
colors agree well with various existing $T_{\rm eff}$--color
relations at $\MoH=-1.0$, with typical differences being well
within $\Delta T_{\rm eff}\sim100$\,K. However, effective
temperatures predicted by the theoretical scales tend to be higher
than those resulting from the empirical relations, by up to
$\sim100$\,K. Similar trends are seen at $\MoH=-2.0$ too,
especially in the $T_{\rm eff}$--$(B-V)$ and $T_{\rm
eff}$--$(V-K)$ planes, where $T_{\rm eff}$ inferred from the
empirical relations are again by about $\sim100$\,K lower than
those predicted by the theoretical scales.

The agreement between the different color--color relations is good
in all color--color planes, both at $\MoH=-1.0$ and $\MoH=-2.0$.
The typical differences are within $\pm0.05$\,mag, although in
certain color--color planes the agreement is even better, to
$\pm0.03$\,mag (e.g., in the $(B-V)$--$(V-I)$ and $(V-I)$--$(V-K)$
planes, both at $\MoH=-1.0$ and $-2.0$).

Finally, there is a good consistency in synthetic colors
calculated with different stellar model atmospheres in $T_{\rm
eff}$--color and color--color planes, both at $\MoH=-1.0$ and
$\MoH=-2.0$. The typical differences are within $\Delta T_{\rm
eff}\sim70$\,K at $\MoH=-1.0$ and $\Delta T_{\rm eff}\sim40$\,K at
$\MoH=-2.0$ in $T_{\rm eff}$--color planes, and within
$\Delta\sim0.07$\,mag at both metallicities in all color--color
planes. Let us note, however, that the differences in the setup of
stellar atmosphere models (opacities, equation of state, etc)
generally become less important at low metallicities which leads
to a better agreement between colors calculated with different
model atmospheres (the differences in photometric colors are
considerably larger at Solar metallicity; see Paper~I).

\acknowledgements

We are grateful to Bertrand Plez (GRAAL, Universit\'{e}
Montpellier) for calculating {\tt MARCS} grid of synthetic
spectra, and numerous comments and discussions. We also thank
Glenn Wahlgren (Lund Observatory) for a careful reading of the
manuscript and his valuable comments and suggestions. AK
acknowledges support from the Wenner-Gren Foundations. This work
was supported in part by grant-in-aids for Scientific Research (C)
and for International Scientific Research (Joint Research) from
the Ministry of Education, Science, Sports and Culture in Japan,
and by a Grant of the Lithuanian State Science and Studies
Foundation. This work was also
supported in part by the P\^ole Scientifique de Mod\'elisation
Num\'erique at ENS-Lyon. Some of the calculations presented in
this paper were performed on the IBM pSeries 690 of the
Norddeutscher Verbund f\"ur Hoch- und H\"ochstleistungsrechnen
(HLRN), and on the IBM SP ``seaborg'' of the NERSC, with support from
the DoE.  We thank all these institutions for a
generous allocation of computer time. This research has also made
use of the SIMBAD and VizieR databases, operated by the CDS,
Strasbourg, France.

\bibliographystyle{aa}

\end{document}